\newif\ifsubmode
\submodefalse

\newif\ifprintfig
\printfigtrue

\newif\ifemulate
\emulatetrue

\ifsubmode
  \documentclass[12pt,preprint]{aastex}
  \received{}
  \accepted{}
  \journalid{}{}
  \articleid{}{}
\else
   \documentclass{emulateapj}
   \submitted{{\it To be submitted for publication in AJ}}
\fi

\renewcommand{\thefootnote}{\arabic{footnote}}

\def\lesssim{\mathrel{\hbox{\rlap{\hbox{\lower4pt\hbox{$\sim$}}}\hbox{$<$}}}}
\def\gtrsim{\mathrel{\hbox{\rlap{\hbox{\lower4pt\hbox{$\sim$}}}\hbox{$>$}}}}

\def\spose#1{\hbox to 0pt{#1\hss}}
\def\simlt{\mathrel{\spose{\lower 3pt\hbox{$\mathchar"218$}}
     \raise 2.0pt\hbox{$\mathchar"13C$}}}
\def\simgt{\mathrel{\spose{\lower 3pt\hbox{$\mathchar"218$}}
     \raise 2.0pt\hbox{$\mathchar"13E$}}}

\slugcomment{Accepted for publication in the Astronomical Journal on August 7, 2013}

\shorttitle{RR Lyrae stars in Segue 2 and 3}
\shortauthors{Boettcher et al.}

\begin{document}

\title{A Search for RR Lyrae Stars in Segue 2 and Segue 3}

\author{Erin Boettcher\altaffilmark{1,2}, Beth Willman\altaffilmark{2}, Ross Fadely\altaffilmark{2,3}, Jay Strader\altaffilmark{4}, Mariah Baker\altaffilmark{2}, Erica Hopkins\altaffilmark{2}, Tonima Tasnim Ananna\altaffilmark{5}, Emily C. Cunningham\altaffilmark{2}, Tim Douglas\altaffilmark{2,6}, Jacob Gilbert\altaffilmark{2}, Annie Preston\altaffilmark{2}, Andrew P. Sturner\altaffilmark{2,7}}

\altaffiltext{1}{Department of Astronomy, University of Wisconsin - Madison, Madison, WI 53706, USA, \texttt{boettche@astro.wisc.edu}}
\altaffiltext{2}{Department of Astronomy, Haverford College, Haverford, PA 19041, USA, \texttt{bwillman@haverford.edu}}
\altaffiltext{3}{Center for Cosmology and Particle Physics, Department of Physics, New York University, New York, NY 10003, USA}
\altaffiltext{4}{Department of Physics and Astronomy, Michigan State University, East Lansing, MI 48824, USA}
\altaffiltext{5}{Department of Physics, Bryn Mawr College, Bryn Mawr, PA 19010, USA}
\altaffiltext{6}{Dropbox Inc., 185 Berry Street, Suite 400, San Francisco, CA 94107, USA}
\altaffiltext{7}{Laboratory for Atmospheric and Space Physics, University of Colorado, Boulder, CO 80303, USA}


\ifsubmode\else
  \ifemulate\else
     \clearpage
  \fi
\fi


\ifsubmode\else
  \ifemulate\else
     \baselineskip=14pt
  \fi
\fi

\begin{abstract}
  We present an extensive search for RR Lyrae stars in and around the
  ultra-faint Milky Way companions Segue 2 and Segue 3. The former
  ($M_{V} = -2.5$, \citealt{Belokurov2009}) appears to be an extremely
  faint dwarf galaxy companion of the Milky Way.  The latter ($M_{V} =
  0.0$, \citealt{Fadely2011}) is among the faintest star clusters known.
  We use $B$ and $V$ band time-series imaging obtained at the WIYN 0.9
  meter telescope at Kitt Peak National Observatory to search for RR
  Lyrae in these objects. In our Segue 2 observations, we present a
  previously unknown fundamental mode (RRab) RR Lyrae star with a
  period of $P_{ab} = 0.748$ days. With this measurement, we revisit
  the inverse correlation between $\langle P_{ab} \rangle$ and
  $\langle [Fe/H] \rangle$ established in the literature for Milky Way
  dwarf galaxies and their RR Lyrae. In this context, the long period
  of Segue 2's RRab star as well as the known significant spread in
  metallicity in this dwarf galaxy are consistent with the observed
  trend in $\langle P_{ab} \rangle$ and $\langle [Fe/H] \rangle$. We
  derive the first robust distance to Segue 2, using both its RRab
  star and spectroscopically confirmed blue horizontal branch
  stars. Using $[Fe/H] = -2.16$ and $-2.44$ dex, we find $d_{RRL} =
  36.6^{+2.5}_{-2.4}$ and $37.7^{+2.7}_{-2.7}$ kpc; assuming $[Fe/H] =
  -2.257$ dex, we find $d_{BHB} = 34.4 \pm 2.6$ kpc. Although no RR
  Lyrae were present in the Segue 3 field, we found a candidate
  eclipsing binary star system.

\end{abstract}

\keywords{galaxies: star clusters ---
          galaxies: dwarf ---
	  stars: distances ---
	  stars: variables: other ---
	  techniques: photometric}

\section{Introduction}\label{intro_sec}
\renewcommand{\thefootnote}{\arabic{footnote}} Over the last decade,
numerous ultra-faint ($M_{V} \gtrsim -8$) companions of the Milky Way
Galaxy have been discovered in Sloan Digital Sky Survey data
\citep[e.g.][]{Belokurov2006b, Belokurov2007, Belokurov2008,
Belokurov2009, Belokurov2010, Koposov2007, Walsh2007, Willman2005a,
Willman2005b, Zucker2006a, Zucker2006b}. Among these discoveries are
the least luminous star clusters known
\citep[e.g.][]{Koposov2007,Belokurov2010,Munoz2012}, as well as the
least luminous, most metal-poor, and most dark matter dominated
galaxies known
\citep[e.g.][]{Kirby2008,Wolf2010,Koposov2011,Simon2011,Willman2011}. Due
to these satellites' low luminosities, it is difficult to determine
their distances, dynamical states, and stellar populations. The use of
RR Lyrae (RRL) stars as standard candles found in time-series
observations has provided an alternative to isochrone fitting for
measuring satellite distances.

RRL stars are short-period (0.3 - 1.0 days, RRab; 0.1 - 0.55 days,
RRc) pulsating variable stars that are found in old and metal-poor
stellar populations \citep{Smith1995, Vivas2004, Sesar2007}. They are
standard candles with mean absolute $V$ band magnitudes of $M_{V} =
0.59 \pm 0.03$ for $[Fe/H] = -1.5$ \citep*{Cacciari2003}. RRL stars
have been found in considerable numbers in all metal-poor components
of the Galaxy; among these RRL are Galactic globular cluster variables
as well as field variables in the halo, thick disk, and bulge. RRL
stars fall into two distinct regions in period/amplitude space and are
thus categorized as fundamental mode (RRab) or first-overtone (RRc)
variables whose light curves exhibit characteristic periods,
amplitudes, and shapes \citep{Smith1995}.  Both types occupy the
intersection of the horizontal branch and the instability strip and
thus range in color from $B - V = 0.18$ to 0.40; additionally,
they display a characteristic increase in $B - V$ at minimum light
\citep{Smith1995}.

The QUEST RR Lyrae survey found that RRab stars exhibit light curves
with mean $V$ band amplitudes of $1.04 \pm 0.24$ mags and mean periods
of $0.539 \pm 0.09$ days. RRc stars have light curves with mean $V$
band amplitudes of $0.536 \pm 0.13$ mags and mean periods of $0.335
\pm 0.07$ days. The former have a distinct saw-toothed shape to their
light curves, while the latter have a smoother shape
\citep{Vivas2004}. \citet{Clement2001} and \citet{Miceli2008} report
mean RRL periods for larger samples of RRL stars in Galactic globular
clusters and in the field, respectively. The former studies both RRab
and RRc stars and finds mean periods for these populations of 0.585
days and 0.349 days, respectively. The latter studies RRab stars alone
and reports a mean period of 0.575 days.

Most of the dwarf companions of the Milky Way Galaxy, including many
of the ultra-faint companions, have been searched for RRL
stars. Bo\"{o}tes I, Canes Venatici II, Coma Berenices, Leo IV, and
Ursa Major II are among the ultra-faint companions known to host one
or more RRL stars \citep{Siegel2006, Dall'Ora2006, Kuehn2008, Greco2008,
  Musella2009, Moretti2009, Dall'Ora2012}. Segue 1, the least-luminous
($M_{V} = -1.5$) dwarf galaxy known, has one published RRL star
\citep{Simon2011}.

Segue 2 \citep[$M_{V} = -2.5 \pm 0.3$,][]{Belokurov2009} and Segue 3
\citep[$M_{V} = 0.0 \pm 0.8$,][]{Fadely2011} are two recently
discovered ultra-faint companions of the Milky Way Galaxy (see Table 1
for the properties of these objects).  Segue 2 is classified as a
dwarf galaxy by \citet{Kirby2013}, because of the significant spread
in the [Fe/H] of its constituent stars \citep{Willman2012}. Segue 3 is
among the lowest-luminosity star clusters known \citep{Fadely2011},
and shares similar properties with a few other extremely
low-luminosity star clusters such as Mu\~noz 1
\citep{Munoz2012}. However, its close proximity ($d \approx 17$ kpc)
makes it a particularly strong candidate for studying such an extreme
stellar system. Additionally, tidal disruption of the object is
suggested by the 11 candidate member stars found more than three
half-light radii from the center of the object
\citep{Fadely2011}. Thus, Segue 3 may be a valuable laboratory for
studying the dynamical evolution of such systems.

\begin{deluxetable}{lcc}
\tabletypesize{\scriptsize}
\tablecaption{Properties of Segue 2 and 3}
\tablehead{ 
\colhead{} &
\colhead{Segue 2} &
\colhead{Segue 3}
}
\startdata
R.A. (J2000) & $2^{h}19^{m}16^{s}$ & $21^{h}21^{m}31^{s}$\\
Decl. & $20^{\circ}10'31''$ & $19^{\circ}07'02''$\\
$M_{V}$ & -2.5 $\pm$ 0.3 & 0.0 $\pm$ 0.8\\
$(m - M)_{0}$ & 17.7 $\pm$ 0.1 & 16.1 $\pm$ 0.1\\
Half-Light Rad. ($r_{H}$) & 3.4$'$ $\pm$ 0.2$'$ & 26$''$ $\pm$ 5$''$\\
\enddata
\tablecomments{Values for Segue 2 are from \citet{Belokurov2009} and values for Segue 3 are from \citet{Fadely2011}.}
\end{deluxetable} 

The accuracy with which the distances to Segue 2 and 3 can be measured
affects the accuracy with which many fundamental physical properties
can be determined. The distance to Segue 2 is currently estimated
using the apparent magnitudes of four candidate blue horizontal branch
members \citep{Belokurov2009}. The distance to Segue 3 is determined
by performing isochrone fitting to spectroscopically selected members
using a maximum likelihood method \citep{Fadely2011}. If one or more
RRL stars can be shown to belong to these objects, then they will
provide a robust complimentary approach to constraining these
distances.

In this paper, we search for RRL stars in and around Segue 2 and 3. In
\S2, we describe the collection and reduction of multi-band
time-series observations at the WIYN 0.9 meter telescope at Kitt Peak
National Observatory. \S3 describes the use of DAOPHOT II and ALLSTAR
II to perform PSF photometry, details the astrometric and photometric
calibration, and describes the selection of variable star
candidates. In \S4, we present a fundamental mode RRL star in Segue 2
and a candidate eclipsing binary star system in the Segue 3 field.  We
use the former as well as three confirmed blue horizontal branch
members to determine the distance to Segue 2 and then consider its
RRL properties in the context of other Milky Way dwarf
galaxies. We conclude with a brief review of our results in \S5.

\section{Data}\label{sec_data}

\subsection{Observations and Data Reduction}\label{ssec_obs}
We obtained Harris $B$ and $V$ band time-series observations of the
Segue 2 and 3 objects using the 0.9 meter WIYN telescope and S2KB CCD
camera at Kitt Peak National Observatory. On October 12th and 13th,
2010, in gray conditions, we obtained 24 (11 $B$, 13 $V$ band) images
of Segue 2 and 42 (20 $B$, 22 $V$ band) images of Segue 3. The seeing
ranged from $1.1''$ to $2.6''$ with a median seeing of $1.7''$. From
October 8th to 11th, 2011, in bright conditions, we took 69 (34 $B$,
35 $V$ band) exposures of Segue 2 and 71 (35 $B$, 36 $V$ band)
exposures of Segue 3. The seeing ranged from $1.2''$ to $3.2''$ with a
median seeing of $1.7''$. For Segue 2, the exposure times ranged from
300 to 600 seconds in $B$ band and 180 to 600 seconds in $V$ band. For
Segue 3, the exposure times varied from 180 to 300 seconds in both $B$
and $V$ band. The exposures were taken alternating between the $B$ and
$V$ bands, and the minimum time between subsequent exposures was set
by a read-out time of approximately three minutes. The S2KB CCD camera
is an array of 2048 by 2048 pixels with a scale of 0.6 arcseconds per
pixel. Both Segue 2 ($r_{H} = 3.4' \pm 0.2'$, \citealt{Belokurov2009})
and Segue 3 ($r_{H} = 0.43' \pm 0.08'$, \citealt{Fadely2011}) were
fully captured within the $20'$ by $20'$ field of view.

To prepare the images for analysis, the exposures were
bias-subtracted, flat-fielded using dome flats, and trimmed to remove
the overscan region. The DAOPHOT II and ALLSTAR II packages were used
to perform PSF photometry on all of the images \citep{Stetson1987,
Stetson1994}. We allowed the point-spread function to vary
quadratically as a function of position. To assess the point source
detection completeness of our photometry, we matched stars from the
Sloan Digital Sky Survey Data Release 7 (SDSS DR7)
\citep{Abazajian2009} to our detected sources within the footprint of
our observations. For both the Segue 2 and 3 fields, our photometric
catalog includes $100\%$ of SDSS stars brighter than $m_{V,0} \sim
19.5$ mags within our footprint, well below the apparent $V$ band
magnitude of the horizontal branches of these objects (see \S3 and
\S4). SDSS DR7 is $\sim 95\%$ complete to $g$, $r \sim 22.2$ mags\footnote{\url{http://www.sdss.org/dr7/}}, so we infer a similarly high point source detection
completeness for our data. While this comparison with SDSS doesn't
itself account for crowding incompleteness, neither Segue 2 nor Segue
3 are crowded in their central regions. We conclude that it is
unlikely that we have missed an RRL star in either Segue 2 or Segue 3
owing to photometric incompleteness, but we cannot rule out the
possibility with $100\%$ confidence.

\subsection{Astrometric Calibration}\label{ssec_astcal}
The online resource
Astrometry.net\footnote{\url{http://nova.astrometry.net/}} was used to
obtain astrometric headers for each exposure. We accessed
Astrometry.net with a Python script to enable automated processing of
all exposures. We used this astrometry to facilitate cross-matching
sources between the S2KB exposures and also to the SDSS DR7 catalog.
Since the astrometry in the SDSS catalog is more precise than we
could obtain for our KPNO data, all coordinates reported in this paper
come from SDSS DR7.

\subsection{Photometric Calibration}\label{ssec_photcal}
The data were photometrically calibrated to SDSS DR7
photometry.\footnote{Downloaded from
\url{http://casjobs.sdss.org/CasJobs/}} We transformed the SDSS $g$
and $r$ magnitudes to $B$ and $V$ using the filter transformations of
\citet{Jordi2006}. The errors are on the order of a few hundredths
of a magnitude for these filter transformations. In the Segue 2 field,
the median errors on our standard magnitudes were thus increased from
$\sim$ 0.018 and 0.017 mag in $g$ and $r$ to $\sim$ 0.037 and 0.020
mag in $B$ and $V$, respectively. For the Segue 3 field, the
corresponding values were $\sim$ 0.012 and 0.011 in $g$ and $r$ and
$\sim$ 0.029 and 0.016 in $B$ and $V$. All magnitudes used for the
light curves and distances presented in this paper have been corrected
for dust using the \citet[][hereafter SFD98]{Schlegel1998} maps
(assuming $R_V$ = 3.1), and using the updated reddening coefficients
presented in \citet{Schlafly2011}. $E(B-V)$ at the center of Segue 2
is 0.183 and at the center of Segue 3 is 0.099.

We used a maximum likelihood analysis to calibrate our data to SDSS as
a function of color, $x$-pixel position, and $y$-pixel position. We
maximize the log-likelihood:

\begin{equation}
\ln\mathcal{L} = -\frac{1}{2}\displaystyle\sum_{i}\displaystyle\sum_{j}\left(\left[\frac{(m_{SDSS,j} - m_{mod,ij})^2}{\sigma_{ij}^2}\right] + \ln{\sigma_{ij}^2}\right)
\end{equation}

where:

\begin{equation}
m_{mod,ij} = m_{instr,ij} + \alpha(B - V)_{SDSS,j} + \beta_{i}x_{i,j} + \gamma_{i}y_{i,j} + \zeta_{i}
\end{equation}

and:

\begin{equation}
\sigma_{i,j}^2 = \sigma_{SDSS,j}^2 + \sigma_{instr,ij}^2 + (\alpha\sigma_{B-V,j})^2
\end{equation}

Here, $i$ refers to the image number and $j$ to the star number.
$m_{instr,ij}$ and $\sigma_{instr,ij}$ are the instrumental magnitudes
and random uncertainties, $m_{SDSS,j}$ and $\sigma_{SDSS,j}$ are the
SDSS magnitudes and uncertainties, $(B - V)_{SDSS, j}$ and
$\sigma_{B-V,j}$ are the SDSS colors and uncertainties, and $x_{i,j}$
and $y_{i,j}$ are $x$- and $y$-coordinates in pixels.  For
$\sigma_{instr,ij}$ we use the value reported by the ALLSTAR II
software.  Thus, a single color term ($\alpha$) is found for each data
set, while a unique $x$ term ($\beta_{i}$), $y$ term ($\gamma_{i}$),
and zero point ($\zeta_{i}$) is found for each exposure. The final
$\alpha$, $\beta_{i}$, $\gamma_{i}$, and $\zeta_{i}$ terms and
corresponding uncertainties are taken to be the median and standard
deviation of the distributions obtained using a bootstrapping
technique. Representative values of $\alpha$, $\beta$, $\gamma$, and
$\zeta$ are shown in Table 2. We apply these terms and propagate their
uncertainties into the calibrated data and corresponding uncertainties
presented in the remainder of this paper.  After the data are
calibrated, evidence of small spatially dependent residuals is present
in the Segue 3 field (but not in the Segue 2 field).  While the
variable star in the Segue 3 field presented in \S4 resides on a part
of the chip that appears relatively unaffected, the photometry for the
star as presented in Table 3 has an additional systematic uncertainty
of a few hundredths of a magnitude.

\begin{deluxetable}{lcc}
\tabletypesize{\scriptsize}
\tablecaption{Photometric Calibration Coefficients}
\tablehead{ 
\colhead{} &
\colhead{Segue 2} &
\colhead{Segue 3}
}
\startdata
$\alpha_{B}$ & 0.113 & 0.160\\
$\beta_{B}$ & $-4.13 \times 10^{-5}$ & $-6.47 \times 10^{-5}$\\
$\gamma_{B}$ & $7.25 \times 10^{-5}$ & $6.16 \times 10^{-5}$\\
$\zeta_{B}$ & 2.77 & 2.71\\
$\alpha_{V}$ & $-0.0620$ & $-0.0505$\\
$\beta_{V}$ & $-3.52 \times 10^{-5}$ & $-4.45 \times 10^{-5}$\\
$\gamma_{V}$ & $9.39 \times 10^{-5}$ & $4.47 \times 10^{-5}$\\
$\zeta_{V}$ & 3.13 & 3.04\\ 
\enddata
\tablecomments{Representative values of the photometric calibration coefficients from Equation (2). The $\beta$, $\gamma$, and $\zeta$ terms are medians among all images for each object and filter.}
\end{deluxetable}

\subsection{Selection of Variable Stars}\label{sec_var}
To select a set of variable star candidates, we quantify the change in
magnitude between the $i$th observation of a given source and the
error-weighted, sigma-clipped average magnitude of the source as:
\begin{equation}
\delta_{mag,i} = \left|\frac{\langle m \rangle - m_{i}}{\sqrt{\sigma_{\langle m \rangle}^{2} + \sigma_{i}^{2}}} \right|
\end{equation}
where $\langle m \rangle$ and $m_{i}$ are the average magnitude and
the $i$th observed magnitude of the source and $\sigma_{\langle m
\rangle}$ and $\sigma_{i}$ are the uncertainties on these
quantities. These include both random and systematic components of the
uncertainty. To obtain an initial set of variable star candidates, we
selected those stars for which $\delta_{mag,i} \ge 3.0$ for at least
three exposures in either passband and from any observing epoch, as
well as those sources which showed a change in magnitude of greater
than 0.5 mags.

We evaluate our ability to identify RRL stars as variable star
candidates using these selection criteria by simulating RRL light
curves at the cadence and precision of our calibrated observations
over the full range of RRL period and amplitude parameter space. We
base our simulations on a representative set of twenty RRab and two
RRc $g$ and $r$ templates provided by \citet{Sesar2010} and
transformed to $B$ and $V$ using the transformations of
\citet{Jordi2006}. The uncertainties are simulated as Gaussian random
errors equal to the observed uncertainties (including both random and
systematic components) for the RRab star in Segue 2 (see \S3), and for
the candidate eclipsing binary in the Segue 3 field (see \S4). We use
the eclipsing binary for Segue 3 because its color and magnitude are
broadly consistent with the instability strip of Segue 3. A total of
45,000 RRab light curves and 2,000 RRc light curves were simulated by
taking linear steps through period, amplitude, and initial phase of
observation. The period was varied from 0.3 to 1.0 days for RRab stars
and from 0.1 to 0.55 days for RRc stars \citep{Vivas2004,
Sesar2007}. The amplitude was varied from 0.4 to 1.8 mags and from 0.1
to 1.0 mags for RRab and RRc stars, respectively \citep{Vivas2004,
Sesar2007}.

We define the detection efficiency of our study as our ability to
select our simulated RRL stars as variable star candidates using the
selection criteria described above. Note that this definition of the
detection efficiency only assesses our ability to flag a source as a
variable star candidate for further evaluation. It does not aim to
assess the accuracy with which we recover the source's input
parameters (period, amplitude, etc.) through means such as light curve
template fitting. Thus, our detection efficiency mainly assesses
whether the cadence of our observations allows for sufficient sampling
of RRL light curves to be able to use the selection criteria defined
above to identify these light curves as variable.

In Segue 3, nearly all of our simulated RRL were identified as
variable star candidates for all periods and amplitudes simulated. In
Segue 2, our identification of variable stars was similarly
successful; however, in several areas of period/amplitude parameter
space, $\gtrsim 10\%$ of sources failed to meet our selection
criteria. These areas included RRab stars with the lowest amplitudes
and longest periods (0.9 - 1.0 days), as well as those with amplitudes
of $\sim 0.4$ mag. They also included RRc stars with amplitudes $<
0.2$ mag, with the longest periods again being the least
detectable. Given that RRab stars with periods greater than 0.9 days
are rare \citep{Miceli2008, Vivas2004}, as are RRc stars with
amplitudes below 0.2 mag
\citep{Vivas2004}, we are confident that the cadence of our
observations combined with our selection criteria are sufficient for
successfully identifying RRL stars as variable star candidates.

In our Segue 2 and 3 observations, approximately 3\% and 2\% of stars were
selected as variable star candidates, respectively. The raw light curves
of these candidate variable stars (magnitude vs. HJD) were then
visually inspected with particular attention paid to stars that varied
in both $B$ and $V$ band, varied throughout the observing period,
and/or showed potentially periodic changes in magnitude. The vast
majority of variable star candidates were falsely identified as
variable due to outliers in otherwise flat light curves, significant
scatter in faint light curves, or proximity to other sources or to the
edges of the exposures. A smaller number of stars ($\sim 3$ per field)
showed non-periodic variation over some or all of the observing
period. Although further observation of these stars may inform our
understanding of the stellar populations in and around Segue 2 and 3,
their lack of periodic variation eliminated them from consideration as
RRL stars. Finally, as discussed in Section 4, one clear periodic
variable star was identified in each of the Segue 2 and 3 fields.


\section{A RR Lyrae in Segue 2}\label{sec_seg2}
A periodic variable star was detected in Segue 2 at (RA, Dec) =
($2^{h}19^{m}0^{s}.06$, $20^{\circ}06'35.15''$) at a distance of 1.6
half-light radii from the center of the object. This star was found to
be a spectroscopic member of Segue 2 by \citet{Kirby2013}. The
variable has a period, amplitude, and sawtooth-shaped light curve
consistent with that of a fundamental mode RRL star (RRab).  As shown
in Figure 1, the variable has a color and magnitude consistent with
being an RRL member of Segue 2 based on past measurements of the
distance to Segue 2 (see
\citealt{Belokurov2009}) and KPNO and SDSS photometry of the object.

To deduce the properties of the variable's light curve, we fit our
observed $B$ and $V$ light curves with the RRab and RRc light curve
templates of \citet{Sesar2010}. The set consists of approximately
twenty RRab and two RRc $ugriz$ templates that span a distribution of
light curve shapes derived from observations. The SDSS $g$ and $r$
magnitudes of the templates were transformed to $B$ and $V$ using the
filter transformations of \citet{Jordi2006}. For all 420 RRab
templates, we initially explored period and amplitude parameter space
by conducting gridded searches for a minimum $\chi^2$ fit. As the
template fitting was performed separately in $B$ and $V$ band, we
define the $\chi^2$ value of a template fit as the sum of the $\chi^2$
values of its $B$ and $V$ band fits. We selected eight RRab templates
as providing the most reasonable fits to the data (i.e. $\chi^2
\lesssim 70$). These templates were used to explore the range of
possible periods and amplitudes using {\tt
emcee}\footnote{\url{http://danfm.ca/emcee/}}, a Markov Chain Monte Carlo
(MCMC) routine \citep{foreman-mackey12}.

\begin{figure}
\plotone{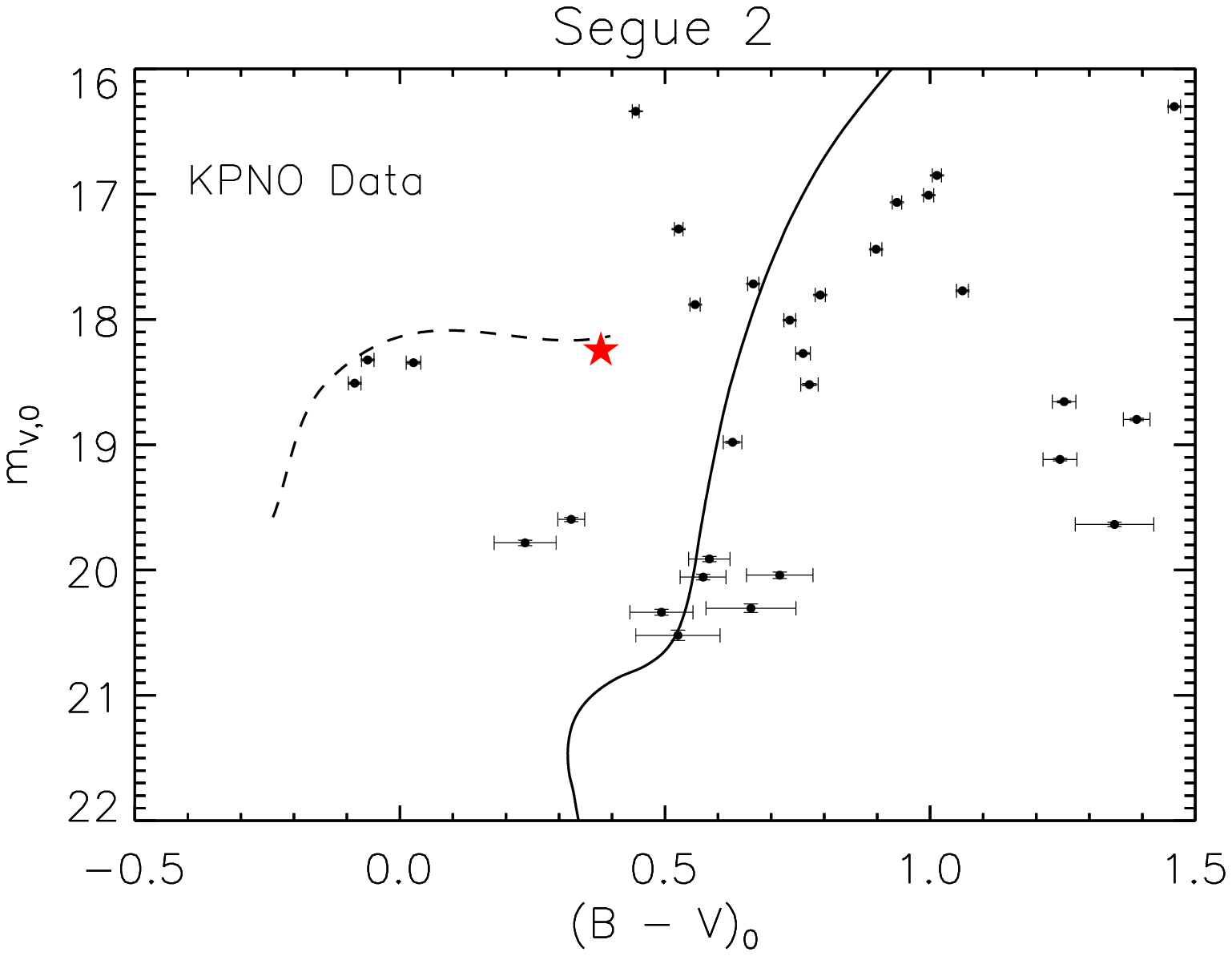}
\plotone{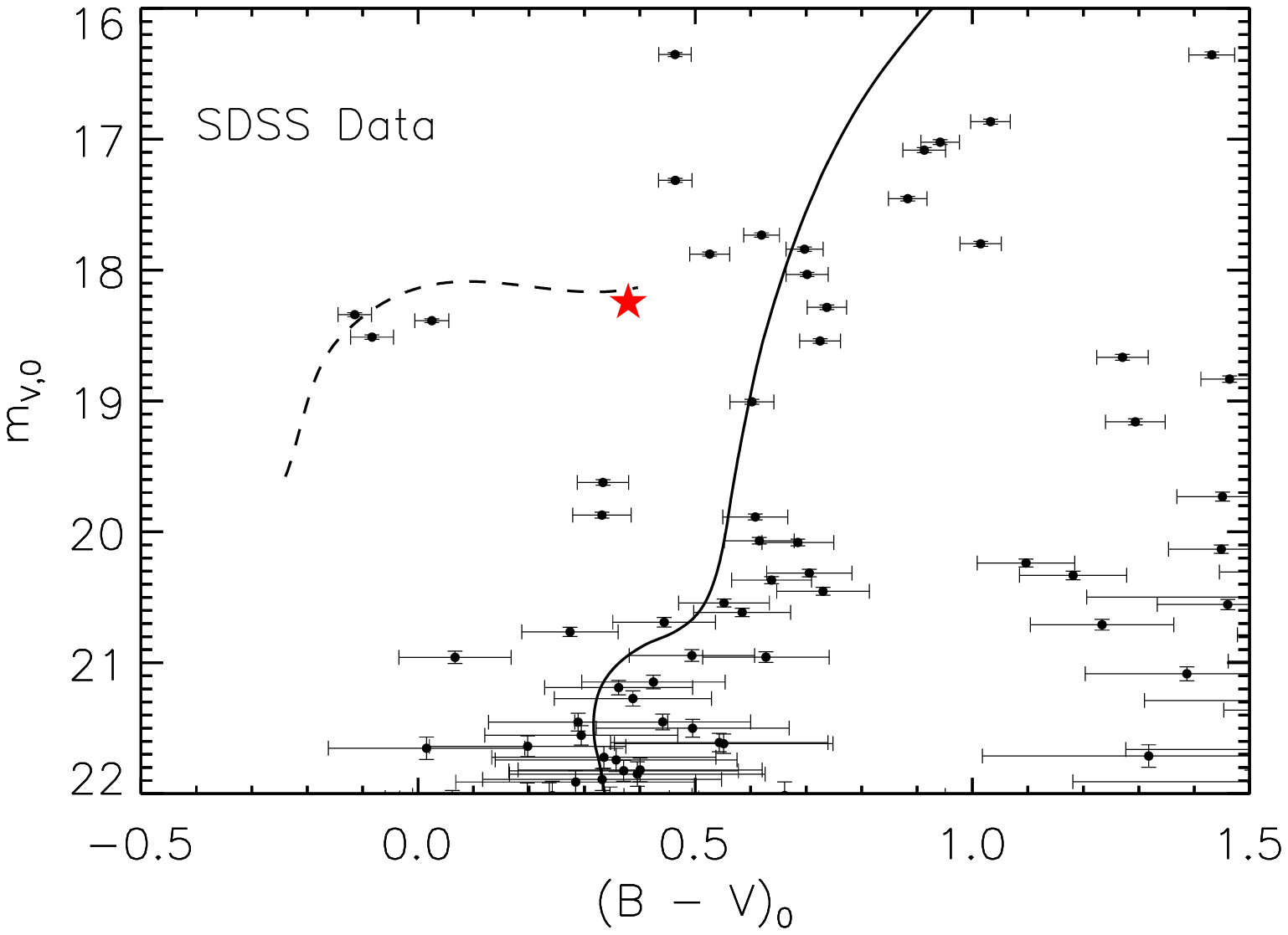}
\caption{The location of the RRab star along the horizontal branch of Segue 2
is shown in KPNO and SDSS photometry of the Segue 2 field within 1
half-light radius of the center of the object. The KPNO photometry of
the variable is indicated by a red star. The isochrone (solid line)
and horizontal branch fiducial (dashed) have $[Fe/H] = -2.257$ and $d =
37.0$ kpc. Although these are overlaid to guide the eye, isochrone
fitting was not performed due to the small number of sources belonging
to the object and the high level of contamination from sources in the
field. The isochrone was obtained from the Dartmouth Stellar Evolution
Database (see \protect\url{http://stellar.dartmouth.edu/models/index.html}),
and the horizontal branch fiducial is a metallicity-corrected combined
M3 and M13 fiducial from \citet{Sand2012}.}
\end{figure}

For each template, the MCMC sampling returns the posterior
distribution of the period and amplitude. We find a bi-modal
distribution of periods and amplitudes consisting of a primary and
secondary peak in the parameter space. Although the template fitting
was performed separately in $B$ and $V$ band, we selected a single
best-fit template having the minimum sum of its $\chi^2$ values in
$B$ and $V$. For the primary peak among RRab templates, this best-fit
template gives a period of $P = 0.748^{+0.006}_{-0.006}$ days, a $B$
band amplitude of $A_{B} = 0.623^{+0.024}_{-0.020}$ mag, and a $V$
band amplitude of $A_{V} = 0.509^{+0.018}_{-0.015}$ mag. At a 68\%
confidence level, the other seven templates are in good agreement with
the first, providing periods as short as 0.742 days and as long as
0.769 days, $B$ band amplitudes that range from 0.596 to 0.667 mag,
and $V$ band amplitudes that range from 0.493 to 0.564 mag.  For the
secondary peak, the MCMC samples imply a shorter period ($P =
0.414^{+0.006}_{-0.002}$ days) and smaller amplitude ($A_{V} =
0.475^{+0.043}_{-0.043}$).  However, the light curve templates in this
peak yield much larger $\chi^2$ values than in the primary peak, and a
visual comparison of the light curves and the data reveals them to
provide a poor match. Furthermore, Lomb-Scargle periodograms
constructed for both bandpasses suggest a most probable period of $P
\sim 0.735$ days, which is broadly consistent with the period of the
primary peak.

Additionally, we assessed the likelihood that the variable could
instead be an RRc star by exploring period and amplitude parameter
space with {\tt emcee} using two RRc templates. However, the RRc
templates generally yield large $\chi^2$ values and poor visual
matches to the data. The only RRc template fits that have $\chi^2$
values comparable to the RRab fits in the primary peak have periods on
the order of 0.75 days, much longer than expected for this class of
variable. Thus, this star does not meet the characteristic profile of
an RRc star.

Figure 2 shows the variable's period folded $B_{0}$, $V_{0}$, and $(B
- V)_{0}$ light curves, with the best fit template overploted.  By
integrating the set of light curve templates consistent with the
observed light curve at a 68\% confidence level or better, we
calculate a flux-averaged $B$ band magnitude of $\langle m_{B}
\rangle_{0} = 18.620^{+0.046}_{-0.021}$ mags and a flux-averaged $V$
band magnitude of $\langle m_{V} \rangle_{0} =
18.246^{+0.032}_{-0.020}$ mags. While using this approach provides a
reasonable estimate of the uncertainty in $\langle m_{B} \rangle_{0}$
and $\langle m_{V} \rangle_{0}$, the resulting uncertainty bars on the
magnitudes are not themselves formal 68\% confidence intervals.  The
star has a median color of $\langle B - V \rangle_{0} = 0.379 \pm
0.012$ and varies in color from $(B - V)_{0} \approx 0.27$ to $(B -
V)_{0} \approx 0.51$ over the course of the pulsational period,
displaying the increase in $B - V$ at minimum light that is
characteristic of RRL stars \citep{Smith1995}.

\begin{figure}
\plotone{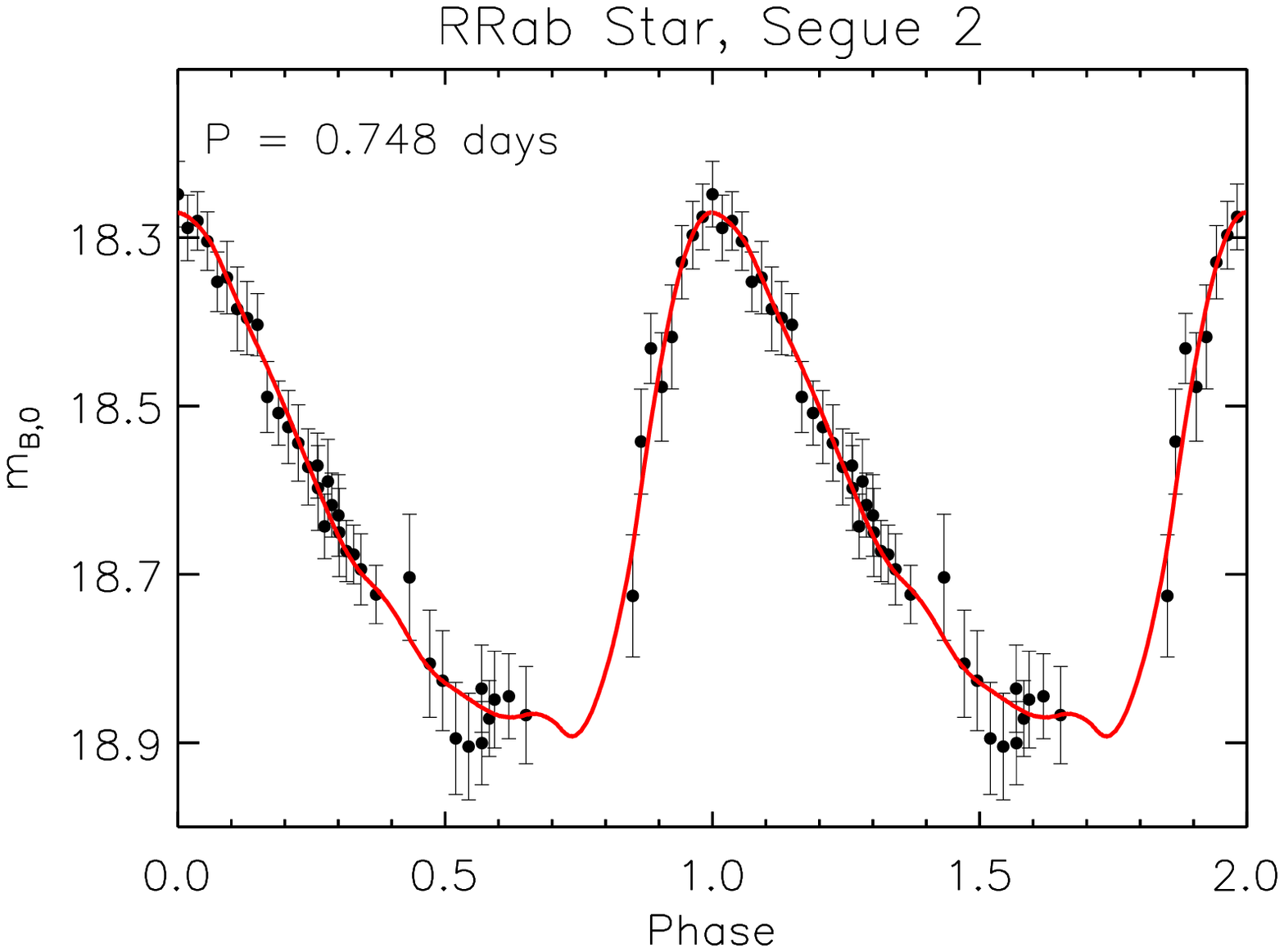}
\plotone{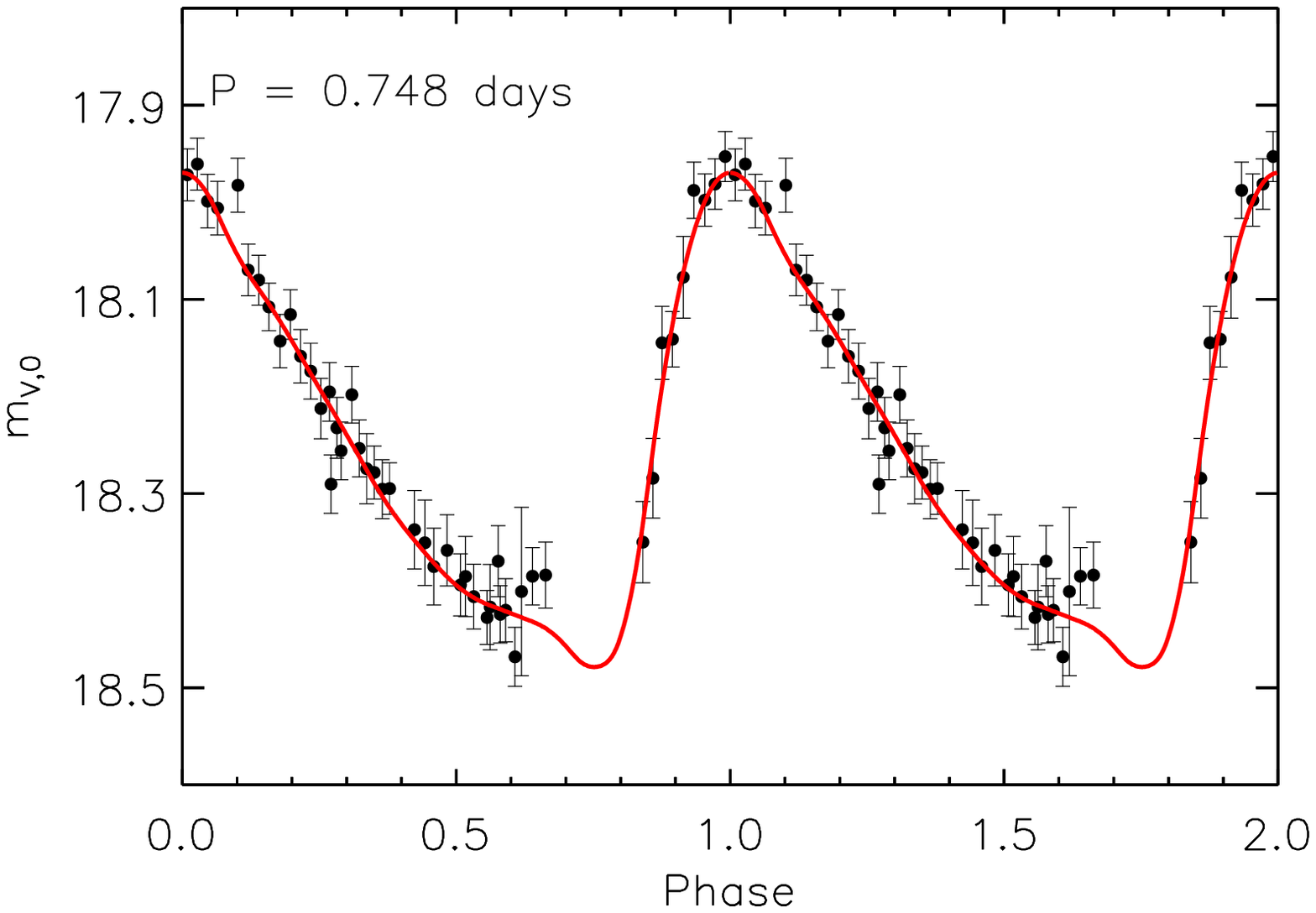}
\plotone{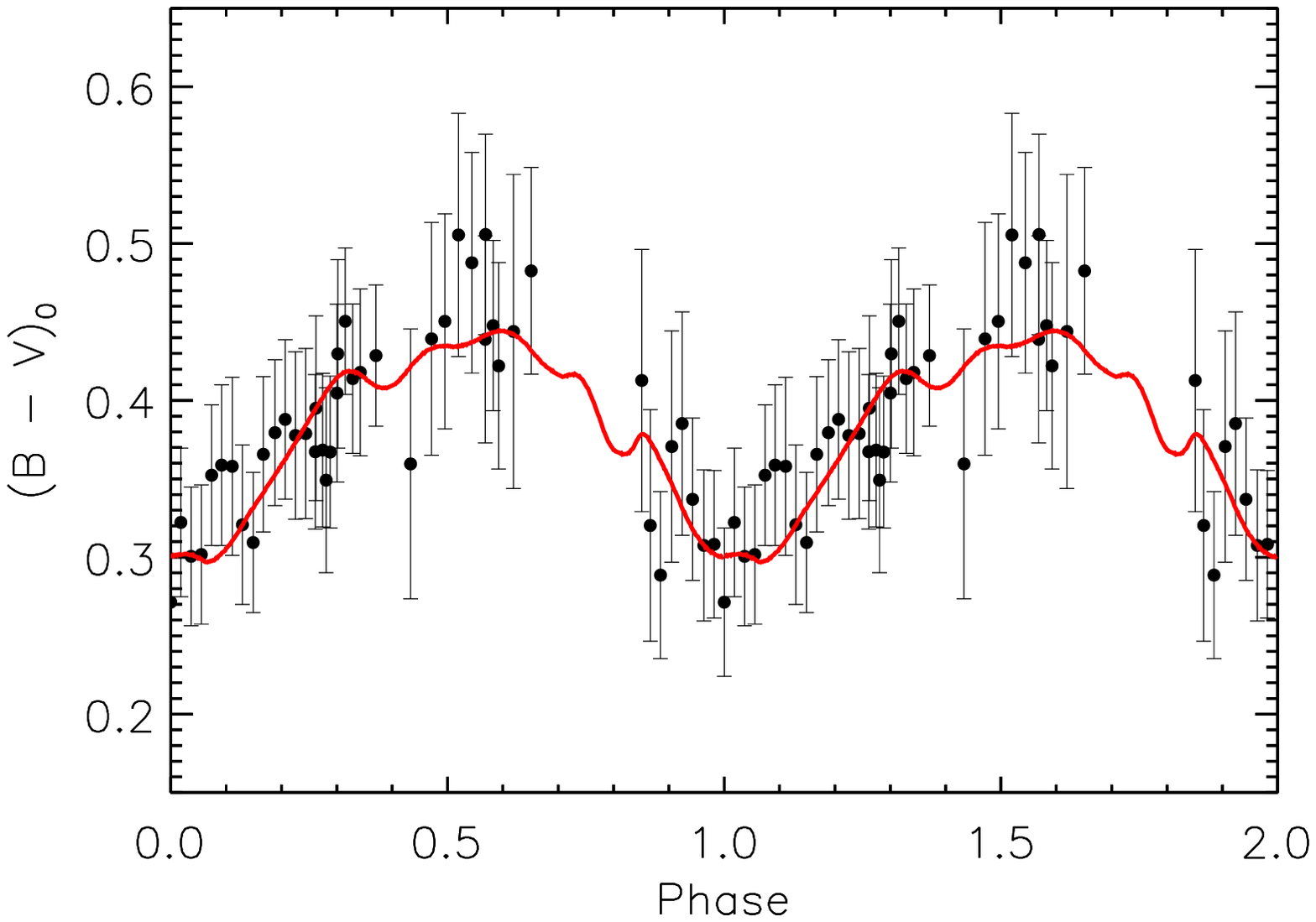}
\caption{The $B_{0}$ and $V_{0}$ light curves of the fundamental
  mode RRL star in Segue 2 are shown in the top and middle
  panels. Overplotted is the best-fit fundamental mode RRL template
  from \citet{Sesar2010}. The $B_{0}$ and $V_{0}$ light curves have
  amplitudes of 0.623 and 0.509 mag, respectively. The $(B - V)_{0}$
  light curve shown in the bottom panel displays the increase in $B -
  V$ at minimum light that is characteristic of RRL stars. The error
  bars include both random and systematic uncertainty.}
\end{figure}

It should be noted that the point-to-point scatter in these light
curves appears smaller than expected given the size of the error bars,
which include both random and systematic uncertainties.  This small
point-to-point scatter suggests that the random uncertainties are
overestimated by ALLSTAR II and/or the systematic uncertainties are
highly correlated from exposure-to-exposure. The systematic
uncertainties account for up to 50\% of the error.  Although the
systematics are likely correlated, we chose to include them in the
error bars because they are derived separately for each exposure and
because they must be included in the error budget for the RRL's
distance estimate in \S3.2.

\subsection{Comparison with the RRL properties of other Milky Way dwarf galaxies}\label{ssec_Oos}

Here, we briefly discuss Segue 2's RRL star in the context of the RRL
populations of other Milky Way dwarf galaxy companions. Historically,
RRab stars in the Milky Way's halo, globular clusters, and dwarf
galaxies have been classified according to their Oosterhoff properties
\citep{Oosterhoff1939}. RRab stars with short periods and large
amplitudes are classified as Oosterhoff I (OoI) stars, and those with
longer periods and smaller amplitudes are deemed Oosterhoff II (OoII)
stars.  Milky Way globular clusters are known to show an Oosterhoff
gap, or an absence of clusters with $0.58 \lesssim \langle
P_{ab}$(days)$\rangle \lesssim 0.62$. However, Milky Way dwarf
galaxies do not display this same dichotomy, and instead largely fall
in the Oosterhoff intermediate to OoII classifications (see, e.g.,
\citealt{Catelan2009}).

Among Galactic globular clusters, an OoII classification is associated
with the most metal-poor systems ([Fe/H] $< -1.5$), with a weak
negative correlation between the mean RRab period and metallicity
\citep{Catelan2009}. For dwarf galaxies of the Milky Way, which are
also metal-poor, a similar correlation
between $\langle P_{ab} \rangle$ and $\langle [Fe/H] \rangle$ also exists
(see, e.g., \citealt{Smith2009, Catelan2009, Clementini2010}).
 
To place the Oosterhoff classification and long period of Segue 2's
RRab star in a more specific context, we revisit the relation between
mean RRab period and mean metallicity for dwarf galaxies using
homogeneously calculated values obtained with a method that remains
robust for ultra-faint systems whose metallicity distributions may be
poorly sampled.  We use the uniformly calculated set of $\langle
[Fe/H] \rangle$ and associated uncertainties found by
\citet{Willman2012} by applying a Bayesian MCMC technique to published
[Fe/H] measurements based on iron lines and their accompanying
uncertainties. A similar technique was applied to re-calculate
$\langle P_{ab} \rangle$ and associated uncertainties from the most
current surveys of variable stars in ultra-faint dwarfs.

In Figure 3, we compare Segue 2's RRab period and mean metallicity
with those of other Milky Way dwarf galaxies with predominantly old
stellar populations. The long period ($P_{ab} = 0.748$ days) and the
amplitude ($A_{V} = 0.509$ mag) of Segue 2's RRab star as well as the
dwarf galaxy's mean metallicity ($\langle [Fe/H] \rangle = -2.257$)
are consistent with an OoII classification (see, e.g.,
\citealt{Kunder2011}). The periods of the single RRL star in each of
Segue 2 and CVn II are longer than the (mean) RRab period of any other
dwarf galaxy. However, the period of the RRab star in Segue 2 and the
mean metallicity of the dwarf galaxy are consistent with the strong
anti-correlation between $\langle P_{ab} \rangle$ and $\langle [Fe/H]
\rangle$ observed in dwarfs given the significant spread in
$[Fe/H]$ confirmed by \citet{Kirby2013}.

We can also compare the RRL populations of Milky Way dwarf galaxy
companions using the RRL specific frequency $S_{RR}$, or the number of
RRL stars per system normalized to a system absolute magnitude of
$M_{V} = -7.5$:
\begin{equation}
S_{RR} = N_{RR}10^{0.4(7.5 + M_{V})}
\end{equation}
Given its one RRL star and an absolute magnitude $M_{V} = -2.5$
\citep{Belokurov2009}, Segue 2 has an RRL specific frequency
$S_{RR} = 100$. Using the absolute magnitudes provided in
\citet{Sand2012}, we find that this is greater than the specific
frequency of any other Milky Way dwarf galaxy companion considered in
Table 4. These objects have $S_{RR}$ values that
range from $\sim 1.3$ (Leo I) to $\sim 60.4$ (ComBer). Of the known
dwarf companions, only Segue 1 ($M_{V} = -1.5$), which is known to
have at least one RRL star \citep{Simon2011}, appears to have a higher
RRL specific frequency. It is necessary to note that as the
completeness of the RRL surveys in these objects is not guaranteed,
the specific frequencies of these objects are lower
limits. Nevertheless, it is interesting to observe that Segue 2 may
have a high specific frequency compared to most other dwarf galaxy
companions in which RRL stars have been discovered.

\begin{figure}
\plotone{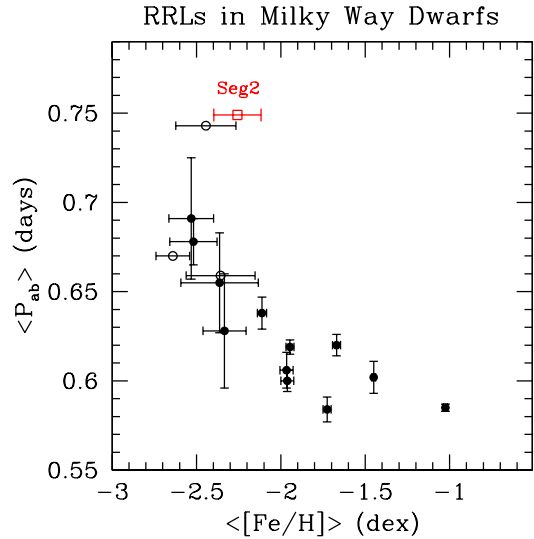}
\caption{Mean RRab period vs. mean [Fe/H] for Milky Way dwarf galaxies
  with predominately old stellar populations.  The error bars are
  uncertainties in the means. The filled circles show galaxies with
  multiple RRab stars; the open circles show objects with either one
  RRab star or for which individual periods are unavailable, so that
  these points have no formal uncertainty in the mean period. The long
  RRab period and mean metallicity of Segue 2 are consistent with the
  established trend in $\langle P_{ab} \rangle$ and $\langle [Fe/H]
  \rangle$ given the significant spread in metallicity established for
  Segue 2 by \citet{Kirby2013}. The data and sources are listed in
  Table 4.}
\end{figure}

\subsection{Distance to Segue 2}\label{ssec_dist}

We calculate the distance to Segue 2 using both its newly identified
RRL star and its blue horizontal branch (BHB) stars (with the
technique described in \S3 of \citealt{Sand2012}). As described
in \S2.3, we used the \citet{Schlafly2011} reddening
coefficients rather than the SFD98 coefficients for the
photometry used to calculate these distances. For both the RRL and the
BHB technique, the updated coefficients make the inferred distance to
Segue 2 several percent larger than it would have been with the SFD98
coefficients.

Using either the RRL star or BHB stars to calculate the distance to
Segue 2 requires knowledge of the stars' metallicities.  Using the
individual metallicities of member stars from \citet{Kirby2013}, we
calculate a mean metallicity of $[Fe/H] = -2.257 \pm 0.140$ (as
described in \S3.1). The BHB [Fe/H] value estimated by
\citet{Belokurov2009} is consistent with this value. Using the three
BHB members and $[Fe/H] = -2.257$, we find a distance to Segue 2 of $d
= 34.4 \pm 2.6$ kpc.

To calculate the RRL distance, we first estimate the metallicity of
the RRL star using each of the following relationships between RRab
metallicity, period, and $V$ band amplitude:
\begin{equation}
[Fe/H] = -8.85[\log P_{ab} + 0.15A_{V}] - 2.60
\end{equation}
\begin{equation}
[Fe/H] = -3.43 - 7.82\log P_{ab}
\end{equation} 
The former relation, from \citet{Alcock2000}, and the latter, from
\citet{Sarajedini2006}, have estimated uncertainties of approximately
0.31 and 0.45 dex, respectively. Using $P_{ab} = 0.748$ days and
$A_{V} = 0.509$ mag, we find $[Fe/H]$ values of $-2.16$ and $-2.44$ dex.

\citet{Chaboyer1999} gives the following relationship between absolute $V$ band magnitude and metallicity for RRL stars:
\begin{equation}
M_{V,RR} = (0.23 \pm 0.04)([Fe/H] + 1.6) + (0.56 \pm 0.12)
\end{equation}
Thus, we derive absolute magnitudes for the RRab star of $M_{V} = 0.43
\pm 0.14$ and $M_{V} = 0.37 \pm 0.15$ for metallicities of $[Fe/H] =
-2.16$ and $[Fe/H] = -2.44$, respectively. Therefore, we find a
distance to the RRab star of $d = 36.6^{+2.5}_{-2.4}$ kpc and $d =
37.7^{+2.7}_{-2.7}$ kpc. All of
our distance measurements individually have $\sim$8\% uncertainty and
are consistent with literature values (see \citealt{Belokurov2009} and
\citealt{Ripepi2012}).

The distances to Segue 2 determined using both the RRL and BHB stars
are consistent between the two techniques within one standard
deviation. Note that this error budget does not include uncertainty in
the metallicity of the RRL and BHB stars, nor the uncertainty in the
absolute value of $E(B-V)$ at the location of the RRL star.  If the
reddening uncertainty is similar to the variation in SFD98 reddening
across the face of Segue 2, this uncertainty may affect the inferred
distance modulus by a couple hundredths of a magnitude (and thus the
distance by $\sim$1\%).

\section{A Candidate Eclipsing Binary in the Segue 3 Field}\label{sec_seg3}
In the Segue 3 field, one periodic variable star was discovered at
(RA, Dec) = ($21^{h}21^{m}41^{s}.21$, $19^{\circ}00'5.51''$) at a
distance of 17 half-light radii from the center of the object. The
variable has a mean $B$ band magnitude of $\langle m_{B} \rangle_{0} =
17.407 \pm 0.018$ mags, a mean $V$ band magnitude of $\langle m_{V}
\rangle_{0} = 16.862 \pm 0.013$ mags, and a mean color of $\langle B -
V \rangle_{0} = 0.550 \pm 0.004$. The star has a likely period of $P
\sim 0.167$ days determined from a Lomb-Scargle periodogram. As shown
in Figure 4, although the variable's apparent brightness is consistent
with the horizontal branch of Segue 3, the star's color is more than
0.1 mag redder in $(B - V)_{0}$ than expected for an RRL
star. Additionally, its light curves are not well fit by the RRc
templates provided by \citet{Sesar2010}. The variable's $m_{B,0}$,
$m_{V,0}$, and $(B - V)_{0}$ light curves are shown in Figure 5.

The light curves of RRc stars and eclipsing binary star systems may
appear deceptively similar in period, amplitude, and shape
\citep*{Kinman2010}. However, while RRL stars vary in $B - V$ over the
course of a pulsational period due to changes in effective
temperature, an eclipsing binary system does not show significant
variation in $B - V$ (see, e.g., Figures 4-6 of
\citealt{Kinman2010}). As shown in Figure 5, the variable star does
not show clear variation in $(B - V)_{0}$ as a function of phase,
suggesting that this variable may be an eclipsing binary.

\begin{figure}
\plotone{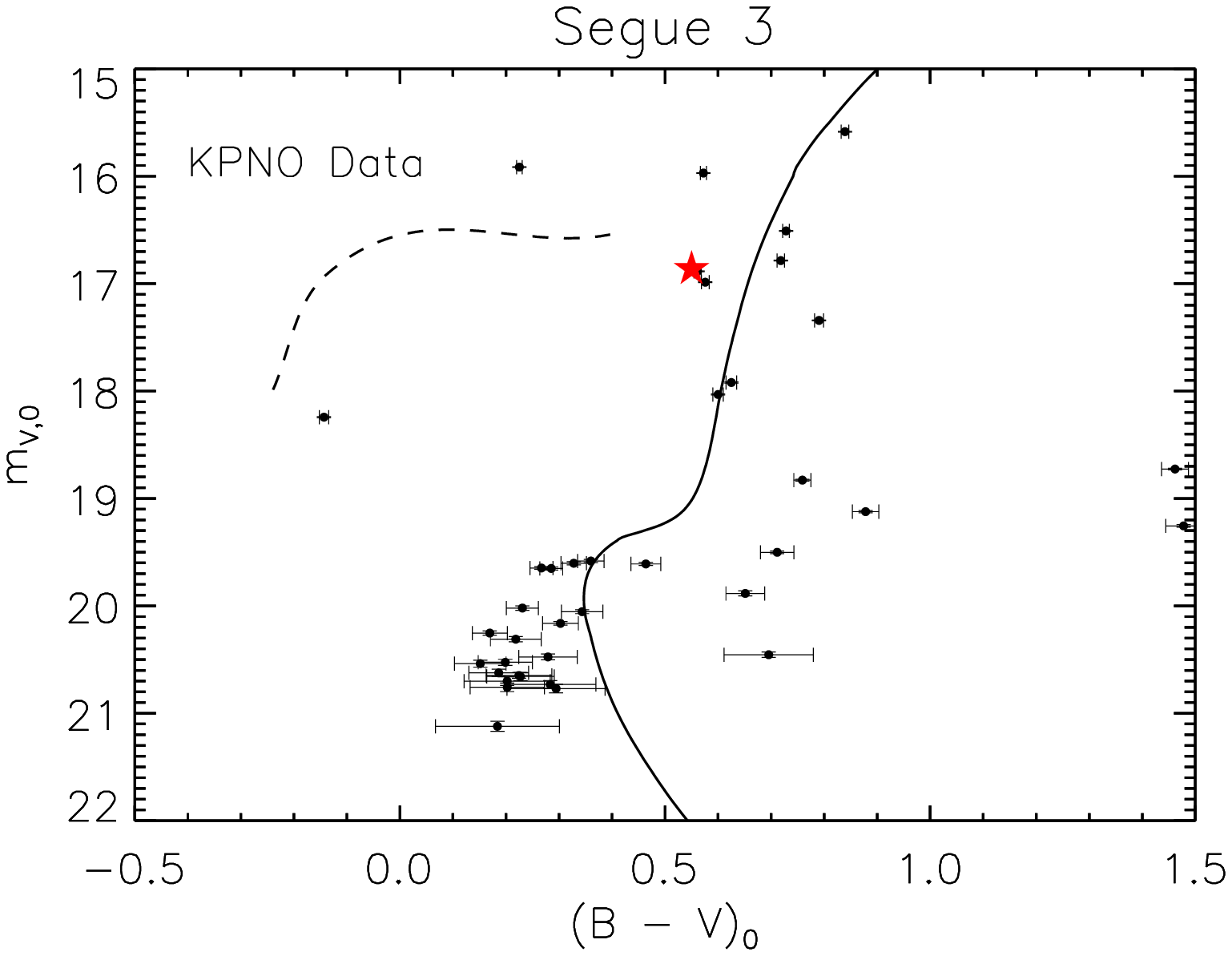}
\plotone{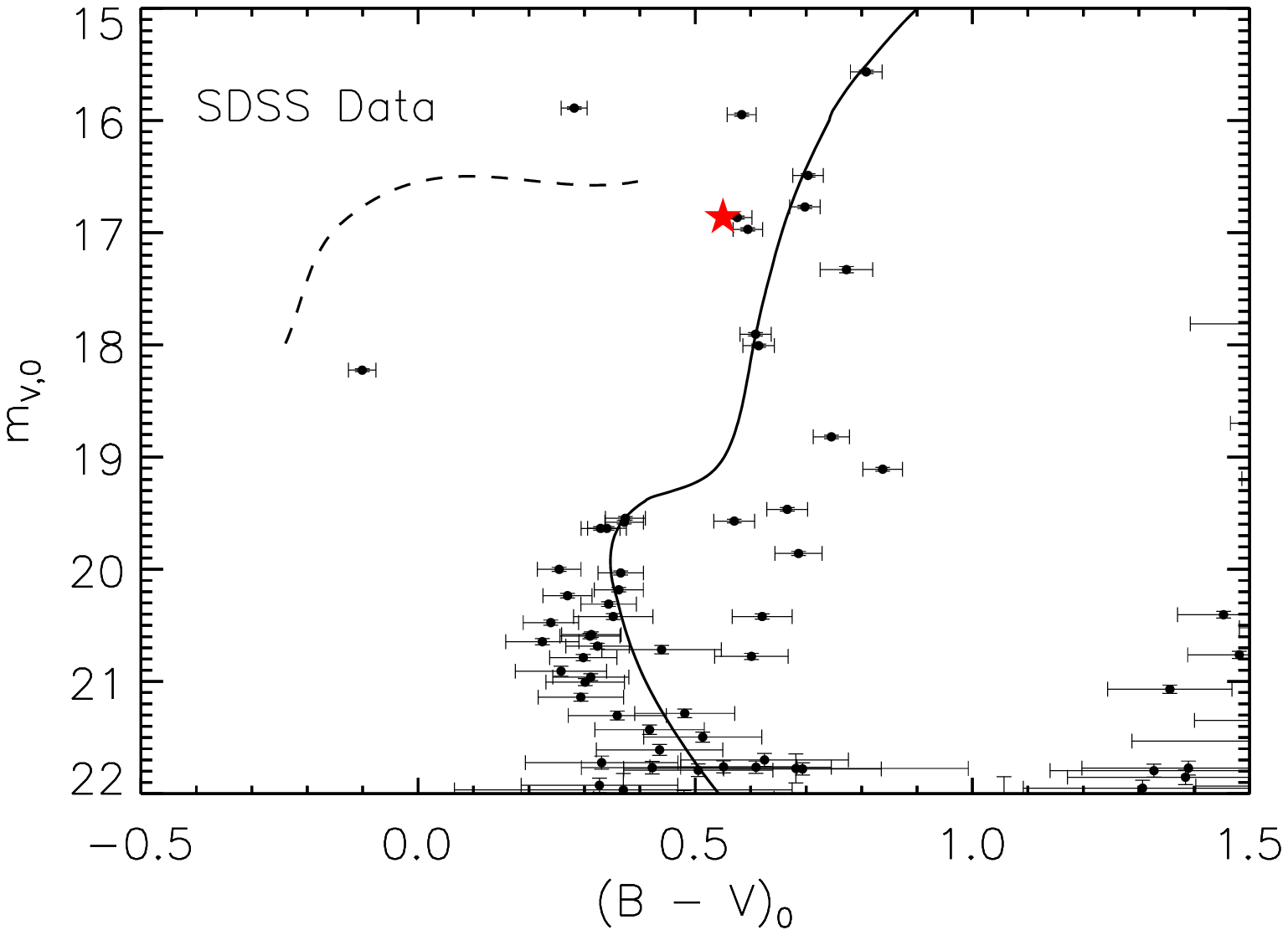}
\caption{The variable star in the Segue 3 field is consistent in magnitude but not in color with the horizontal branch of Segue 3 as indicated by KPNO and SDSS photometry of the object within 3 half-light radii of the center. The KPNO photometry of the variable is indicated by a red star. The isochrone (solid line) and horizontal branch fiducial (dashed) have $[Fe/H] = -1.7$ and $d = 16.9$ kpc and were obtained from the same sources as those in Figure 1.}
\end{figure}

\begin{figure}
\plotone{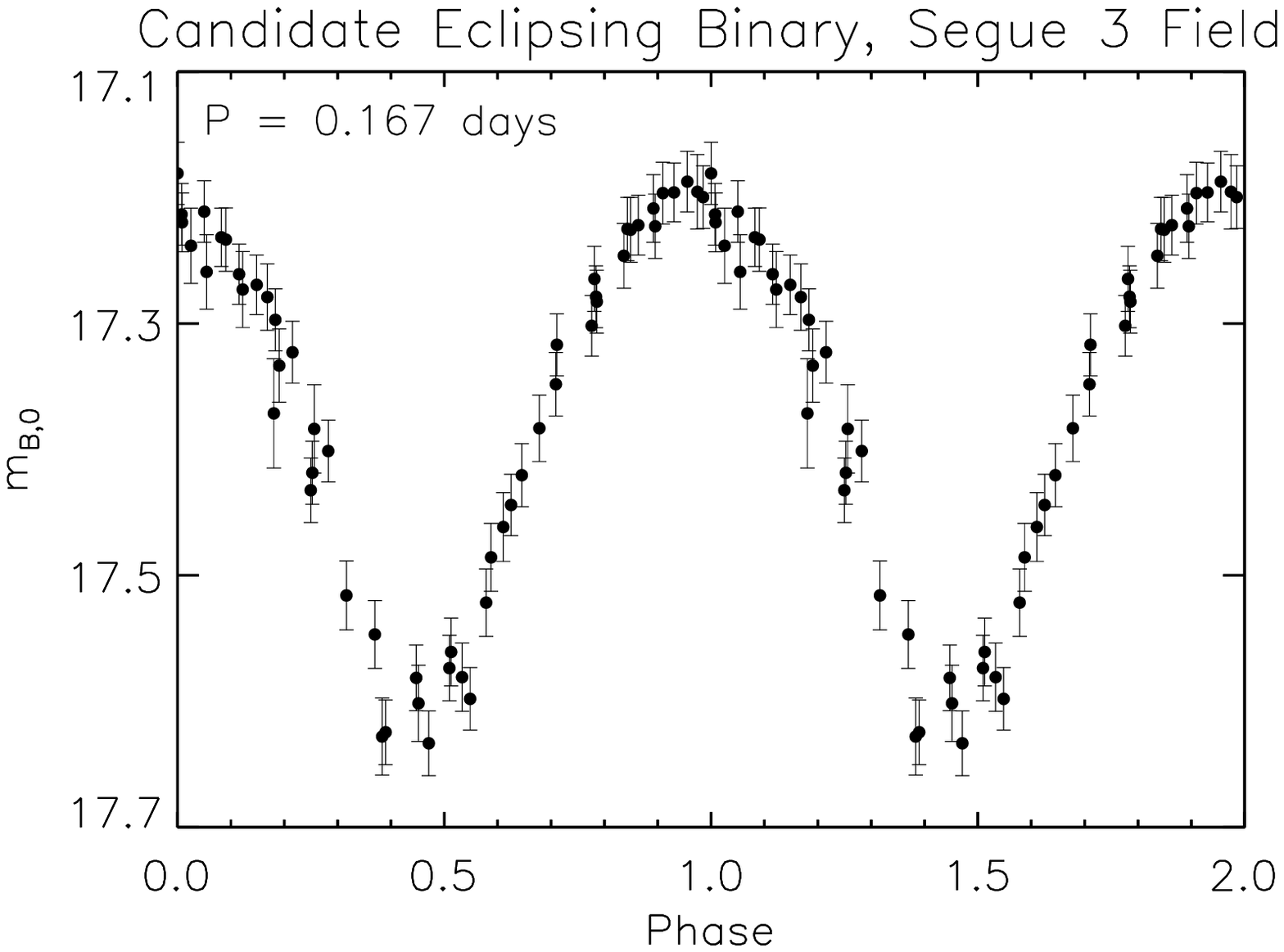}
\plotone{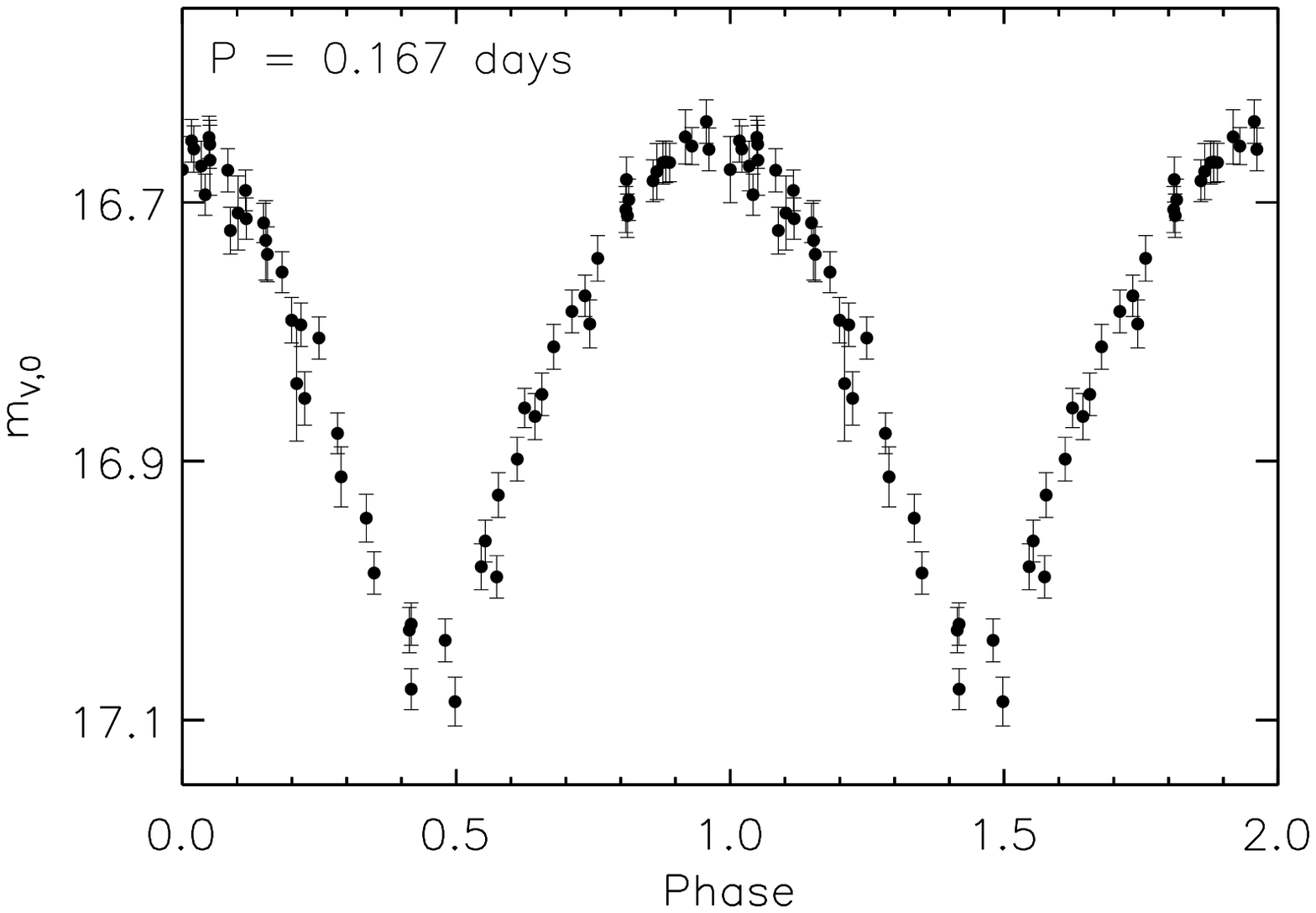}
\plotone{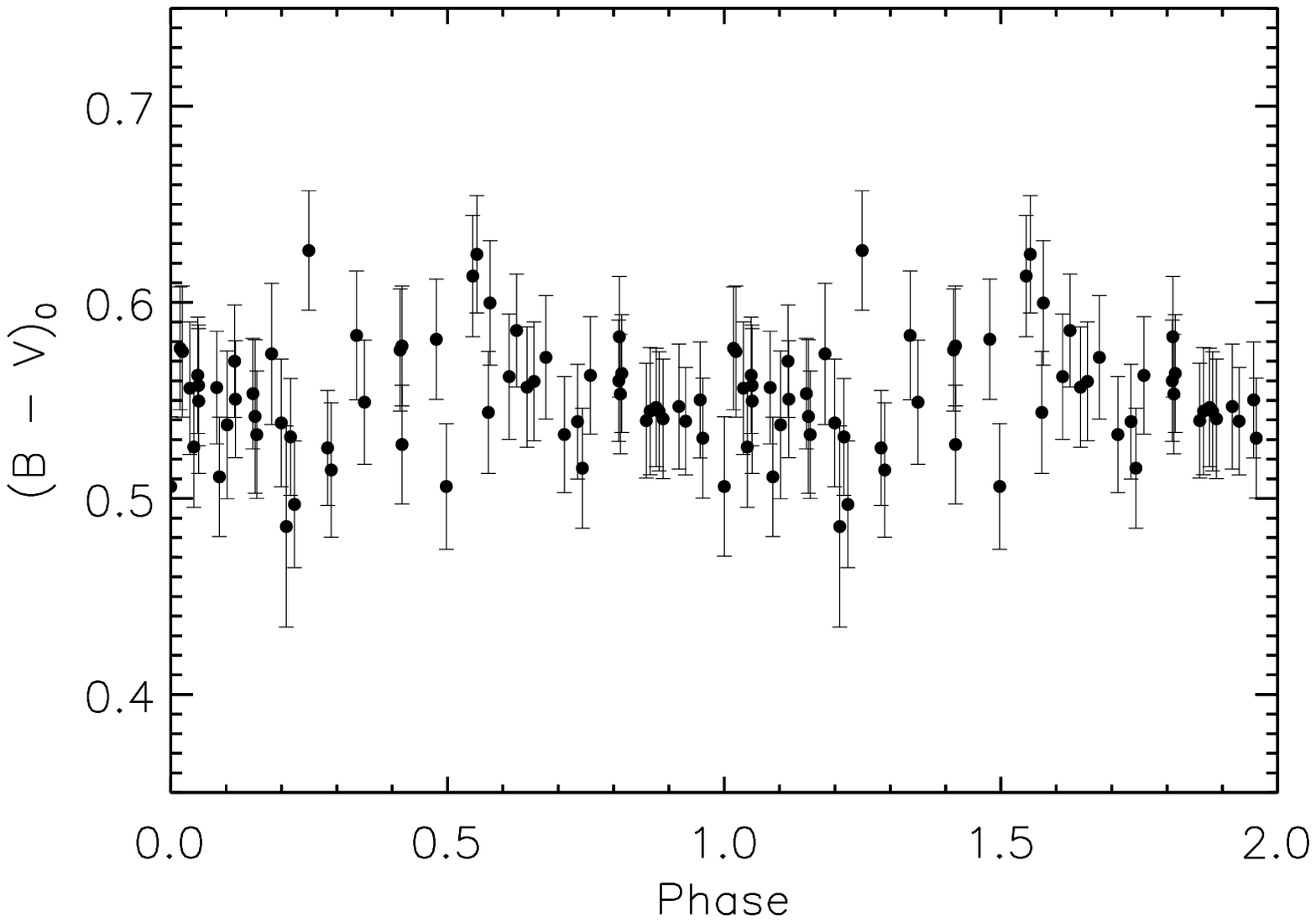}
\caption{The $B_{0}$ and $V_{0}$ light curves of a periodic
  variable star in the Segue 3 field are shown in the top and middle
  panels. These light curves are not well fit by the RRc light curve
  templates of \citet{Sesar2010}, suggesting that this star is not an
  RRL candidate. The $(B - V)_{0}$ light curve shown in the bottom
  panel does not show clear variation as a function of phase and
  supports the hypothesis that this star is an eclipsing binary. The
  error bars include both random and systematic components.}
\end{figure}

\begin{deluxetable}{lcc}
\tabletypesize{\scriptsize}
\tablecaption{}
\tablehead{ 
\colhead{} &
\colhead{Segue 2 Variable} &
\colhead{Segue 3 Variable}
}
\startdata
R.A. (J2000) & $2^{h}19^{m}0^{s}.06$ & $21^{h}21^{m}41^{s}.21$\\
Decl. & $20^{\circ}06'35.15''$ & $19^{\circ}00'5.51''$\\
$\langle m_{V} \rangle_{0}$ & 18.25$^{+0.03}_{-0.02}$ & 16.86 $\pm$ 0.01\\
$\langle B - V \rangle_{0}$ & 0.38 $\pm$ 0.012 & 0.55 $\pm$ 0.004\\
Amplitude ($B$) & 0.62$^{+0.024}_{-0.020}$ & $\sim$0.45\\
Amplitude ($V$) & 0.51$^{+0.018}_{-0.015}$ & $\sim$0.45\\
Period (Days) & 0.748$^{+0.006}_{-0.006}$ & $\sim$0.167\\
$E(B - V)$ & 0.220 & 0.102\\
Classification & RRab & candidate eclipsing binary\\
\enddata
\end{deluxetable}

\section{Conclusions}\label{sec_con}
We have used multi-band time-series photometry obtained at the WIYN
0.9 meter telescope at KPNO to conduct a complete search for RRL stars
in Segue 2 and Segue 3. We have discovered an RRL star with properties
consistent with a fundamental mode RRL (RRab) star in Segue 2, and a
candidate eclipsing binary in the Segue 3 field.  We derive the first
robust distance to Segue 2 using both its RRL star and
spectroscopically confirmed BHB stars. The latter method yields a
distance of $d = 34.4 \pm 2.6$ kpc (for $[Fe/H] = -2.257$), while the
former method gives distances of $d = 36.6^{+2.5}_{-2.4}$ kpc and $d =
37.7^{+2.7}_{-2.7}$ kpc for $[Fe/H] = -2.16$ and $-2.44$,
respectively. These distances are consistent with one another to
within one standard deviation. Future spectroscopic measurements of
the RRab and the BHB stars' [Fe/H] will facilitate an even more robust
measurement of the distance to Segue 2.

We revisit the known anti-correlation between $\langle P_{ab} \rangle$
and $\langle [Fe/H] \rangle$ for RRL in Milky Way dwarf galaxies,
using a uniformly calculated set of $\langle [Fe/H] \rangle$ and
$\langle P_{ab} \rangle$. Placing the 0.748 day period of the Segue 2
RRab star in this context, we find that the RRab period and mean
metallicity of Segue 2 are consistent with the established trend given
the significant spread in metallicity in Segue 2 demonstrated by
\citet{Kirby2013}. The tightness of the observed inverse correlation
between $\langle P_{ab} \rangle$ and $\langle [Fe/H] \rangle$ in dwarf
galaxies is worthy of careful, continued study as more RRL are found
in these objects. This relation may ultimately yield an interesting
avenue for inference of the chemical properties of diffuse streams and
distant ultra-faint dwarfs in the era of time domain surveys such as
LSST.

\acknowledgments

EB, BW, RF, MB, and EH acknowledge support from NSF AST-0908193 and
NSF AST-1151462.  EB, TTA, ECC, TD, JG, AP, and APS acknowledge
observatory travel support from Haverford College's Green Fund.  We
thank Branimir Sesar, Gisella Clementini, Dustin Lang, and Marla Geha
for helpful comments and conversations, and Scott Engle for his help
with data collection.  We thank Joshua Haislip for providing a Python
script for interfacing with Astrometry.net.  We acknowledge
Haverford's Koshland Integrated Natural Sciences Center for supporting
WIYN 0.9m membership dues.  We thank Joe Cammisa for his computing
support and Hillary Mathis for training us to use the WIYN 0.9m
telescope and S2KB camera. We also thank the anonymous referee for
very helpful comments. This work has made use of NASA's Astrophysics
Data System.

\bibliographystyle{apj}

\begin{thebibliography}{63}
\expandafter\ifx\csname natexlab\endcsname\relax\def\natexlab#1{#1}\fi

\bibitem[{{Abazajian} {et~al.}(2009){Abazajian}, {Adelman-McCarthy},
  {Ag{\"u}eros}, {Allam}, {Allende Prieto}, {An}, {Anderson}, {Anderson},
  {Annis}, {Bahcall}, \& et~al.}]{Abazajian2009}
{Abazajian}, K.~N., {Adelman-McCarthy}, J.~K., {Ag{\"u}eros}, M.~A., {et~al.}
  2009, \apjs, 182, 543

\bibitem[{{Alcock} {et~al.}(2000){Alcock}, {Allsman}, {Alves}, {Axelrod},
  {Basu}, {Becker}, {Bennett}, {Cook}, {Drake}, {Freeman}, {Geha}, {Griest},
  {King}, {Lehner}, {Marshall}, {Minniti}, {Nelson}, {Peterson}, {Popowski},
  {Pratt}, {Quinn}, {Stubbs}, {Sutherland}, {Tomaney}, {Vandehei}, \&
  {Welch}}]{Alcock2000}
{Alcock}, C., {Allsman}, R.~A., {Alves}, D.~R., {et~al.} 2000, \aj, 119, 2194

\bibitem[{{Belokurov} {et~al.}(2006){Belokurov}, {Zucker}, {Evans},
  {Wilkinson}, {Irwin}, {Hodgkin}, {Bramich}, {Irwin}, {Gilmore}, {Willman},
  {Vidrih}, {Newberg}, {Wyse}, {Fellhauer}, {Hewett}, {Cole}, {Bell}, {Beers},
  {Rockosi}, {Yanny}, {Grebel}, {Schneider}, {Lupton}, {Barentine},
  {Brewington}, {Brinkmann}, {Harvanek}, {Kleinman}, {Krzesinski}, {Long},
  {Nitta}, {Smith}, \& {Snedden}}]{Belokurov2006b}
{Belokurov}, V., {Zucker}, D.~B., {Evans}, N.~W., {et~al.} 2006, \apjl, 647,
  L111

\bibitem[{{Belokurov} {et~al.}(2007){Belokurov}, {Zucker}, {Evans}, {Kleyna},
  {Koposov}, {Hodgkin}, {Irwin}, {Gilmore}, {Wilkinson}, {Fellhauer},
  {Bramich}, {Hewett}, {Vidrih}, {De Jong}, {Smith}, {Rix}, {Bell}, {Wyse},
  {Newberg}, {Mayeur}, {Yanny}, {Rockosi}, {Gnedin}, {Schneider}, {Beers},
  {Barentine}, {Brewington}, {Brinkmann}, {Harvanek}, {Kleinman}, {Krzesinski},
  {Long}, {Nitta}, \& {Snedden}}]{Belokurov2007}
---. 2007, \apj, 654, 897

\bibitem[{{Belokurov} {et~al.}(2008){Belokurov}, {Walker}, {Evans}, {Faria},
  {Gilmore}, {Irwin}, {Koposov}, {Mateo}, {Olszewski}, \&
  {Zucker}}]{Belokurov2008}
{Belokurov}, V., {Walker}, M.~G., {Evans}, N.~W., {et~al.} 2008, \apjl, 686,
  L83

\bibitem[{{Belokurov} {et~al.}(2009){Belokurov}, {Walker}, {Evans}, {Gilmore},
  {Irwin}, {Mateo}, {Mayer}, {Olszewski}, {Bechtold}, \&
  {Pickering}}]{Belokurov2009}
---. 2009, \mnras, 397, 1748

\bibitem[{{Belokurov} {et~al.}(2010){Belokurov}, {Walker}, {Evans}, {Gilmore},
  {Irwin}, {Just}, {Koposov}, {Mateo}, {Olszewski}, {Watkins}, \&
  {Wyrzykowski}}]{Belokurov2010}
---. 2010, \apjl, 712, L103

\bibitem[{{Bersier} \& {Wood}(2002)}]{Bersier2002}
{Bersier}, D., \& {Wood}, P.~R. 2002, \aj, 123, 840

\bibitem[{{Bonanos} {et~al.}(2004){Bonanos}, {Stanek}, {Szentgyorgyi},
  {Sasselov}, \& {Bakos}}]{Bonanos2004}
{Bonanos}, A.~Z., {Stanek}, K.~Z., {Szentgyorgyi}, A.~H., {Sasselov}, D.~D., \&
  {Bakos}, G.~{\'A}. 2004, \aj, 127, 861

\bibitem[{{Cacciari} \& {Clementini}(2003)}]{Cacciari2003}
{Cacciari}, C., \& {Clementini}, G. 2003, in Lecture Notes in Physics, Berlin
  Springer Verlag, Vol. 635, Stellar Candles for the Extragalactic Distance
  Scale, ed. D.~{Alloin} \& W.~{Gieren}, 105--122

\bibitem[{{Catelan}(2009)}]{Catelan2009}
{Catelan}, M. 2009, \apss, 320, 261

\bibitem[{{Chaboyer}(1999)}]{Chaboyer1999}
{Chaboyer}, B. 1999, Post-Hipparcos Cosmic Candles, 237, 111

\bibitem[{{Clement} {et~al.}(2001){Clement}, {Muzzin}, {Dufton}, {Ponnampalam},
  {Wang}, {Burford}, {Richardson}, {Rosebery}, {Rowe}, \& {Hogg}}]{Clement2001}
{Clement}, C.~M., {Muzzin}, A., {Dufton}, Q., {et~al.} 2001, \aj, 122, 2587

\bibitem[{{Clementini}(2010)}]{Clementini2010}
{Clementini}, G. 2010, in Variable Stars, the Galactic halo and Galaxy
  Formation, ed. C.~{Sterken}, N.~{Samus}, \& L.~{Szabados}, 107

\bibitem[{{Dall'Ora} {et~al.}(2006){Dall'Ora}, {Clementini}, {Kinemuchi},
  {Ripepi}, {Marconi}, {Di Fabrizio}, {Greco}, {Rodgers}, {Kuehn}, \&
  {Smith}}]{Dall'Ora2006}
{Dall'Ora}, M., {Clementini}, G., {Kinemuchi}, K., {et~al.} 2006, \apjl, 653,
  L109

\bibitem[{{Dall'Ora} {et~al.}(2012){Dall'Ora}, {Kinemuchi}, {Ripepi},
  {Rodgers}, {Clementini}, {Di Fabrizio}, {Smith}, {Marconi}, {Musella},
  {Greco}, {Kuehn}, {Catelan}, {Pritzl}, \& {Beers}}]{Dall'Ora2012}
{Dall'Ora}, M., {Kinemuchi}, K., {Ripepi}, V., {et~al.} 2012, \apj, 752, 42

\bibitem[{{Fadely} {et~al.}(2011){Fadely}, {Willman}, {Geha}, {Walsh},
  {Mu{\~n}oz}, {Jerjen}, {Vargas}, \& {Da Costa}}]{Fadely2011}
{Fadely}, R., {Willman}, B., {Geha}, M., {et~al.} 2011, \aj, 142, 88

\bibitem[{{Foreman-Mackey} {et~al.}(2012){Foreman-Mackey}, {Hogg}, {Lang}, \&
  {Goodman}}]{foreman-mackey12}
{Foreman-Mackey}, D., {Hogg}, D.~W., {Lang}, D., \& {Goodman}, J. 2012, ArXiv
  e-prints

\bibitem[{{Garofalo} {et~al.}(2013){Garofalo}, {Cusano}, {Clementini},
  {Ripepi}, {Dall'Ora}, {Moretti}, {Coppola}, {Musella}, \&
  {Marconi}}]{Garofalo2013}
{Garofalo}, A., {Cusano}, F., {Clementini}, G., {et~al.} 2013, \apj, 767, 62

\bibitem[{{Greco} {et~al.}(2008){Greco}, {Dall'Ora}, {Clementini}, {Ripepi},
  {Di Fabrizio}, {Kinemuchi}, {Marconi}, {Musella}, {Smith}, {Rodgers},
  {Kuehn}, {Beers}, {Catelan}, \& {Pritzl}}]{Greco2008}
{Greco}, C., {Dall'Ora}, M., {Clementini}, G., {et~al.} 2008, \apjl, 675, L73

\bibitem[{{Held} {et~al.}(2001){Held}, {Clementini}, {Rizzi}, {Momany},
  {Saviane}, \& {Di Fabrizio}}]{Held2001}
{Held}, E.~V., {Clementini}, G., {Rizzi}, L., {et~al.} 2001, \apjl, 562, L39

\bibitem[{{Jordi} {et~al.}(2006){Jordi}, {Grebel}, \& {Ammon}}]{Jordi2006}
{Jordi}, K., {Grebel}, E.~K., \& {Ammon}, K. 2006, \aap, 460, 339

\bibitem[{{Kaluzny} {et~al.}(1995){Kaluzny}, {Kubiak}, {Szymanski}, {Udalski},
  {Krzeminski}, \& {Mateo}}]{Kaluzny1995}
{Kaluzny}, J., {Kubiak}, M., {Szymanski}, M., {et~al.} 1995, \aaps, 112, 407

\bibitem[{{Kinman} \& {Brown}(2010)}]{Kinman2010}
{Kinman}, T.~D., \& {Brown}, W.~R. 2010, \aj, 139, 2014

\bibitem[{{Kirby} {et~al.}(2013){Kirby}, {Boylan-Kolchin}, {Cohen}, {Geha},
  {Bullock}, \& {Kaplinghat}}]{Kirby2013}
{Kirby}, E.~N., {Boylan-Kolchin}, M., {Cohen}, J.~G., {et~al.} 2013, ArXiv
  e-prints

\bibitem[{{Kirby} {et~al.}(2008){Kirby}, {Simon}, {Geha}, {Guhathakurta}, \&
  {Frebel}}]{Kirby2008}
{Kirby}, E.~N., {Simon}, J.~D., {Geha}, M., {Guhathakurta}, P., \& {Frebel}, A.
  2008, \apjl, 685, L43

\bibitem[{{Kirby} {et~al.}(2010){Kirby}, {Guhathakurta}, {Simon}, {Geha},
  {Rockosi}, {Sneden}, {Cohen}, {Sohn}, {Majewski}, \& {Siegel}}]{Kirby2010}
{Kirby}, E.~N., {Guhathakurta}, P., {Simon}, J.~D., {et~al.} 2010, \apjs, 191,
  352

\bibitem[{{Koposov} {et~al.}(2007){Koposov}, {de Jong}, {Belokurov}, {Rix},
  {Zucker}, {Evans}, {Gilmore}, {Irwin}, \& {Bell}}]{Koposov2007}
{Koposov}, S., {de Jong}, J.~T.~A., {Belokurov}, V., {et~al.} 2007, \apj, 669,
  337

\bibitem[{{Koposov} {et~al.}(2011){Koposov}, {Gilmore}, {Walker}, {Belokurov},
  {Wyn Evans}, {Fellhauer}, {Gieren}, {Geisler}, {Monaco}, {Norris}, {Okamoto},
  {Pe{\~n}arrubia}, {Wilkinson}, {Wyse}, \& {Zucker}}]{Koposov2011}
{Koposov}, S.~E., {Gilmore}, G., {Walker}, M.~G., {et~al.} 2011, \apj, 736, 146

\bibitem[{{Kuehn} {et~al.}(2008){Kuehn}, {Kinemuchi}, {Ripepi}, {Clementini},
  {Dall'Ora}, {Di Fabrizio}, {Rodgers}, {Greco}, {Marconi}, {Musella}, {Smith},
  {Catelan}, {Beers}, \& {Pritzl}}]{Kuehn2008}
{Kuehn}, C., {Kinemuchi}, K., {Ripepi}, V., {et~al.} 2008, \apjl, 674, L81

\bibitem[{{Kunder} {et~al.}(2011){Kunder}, {Walker}, {Stetson}, {Bono},
  {Nemec}, {de Propris}, {Monelli}, {Cassisi}, {Andreuzzi}, {Dall'Ora}, {Di
  Cecco}, \& {Zoccali}}]{Kunder2011}
{Kunder}, A., {Walker}, A., {Stetson}, P.~B., {et~al.} 2011, \aj, 141, 15

\bibitem[{{Mateo} {et~al.}(1995){Mateo}, {Fischer}, \&
  {Krzeminski}}]{Mateo1995}
{Mateo}, M., {Fischer}, P., \& {Krzeminski}, W. 1995, \aj, 110, 2166

\bibitem[{{Miceli} {et~al.}(2008){Miceli}, {Rest}, {Stubbs}, {Hawley}, {Cook},
  {Magnier}, {Krisciunas}, {Bowell}, \& {Koehn}}]{Miceli2008}
{Miceli}, A., {Rest}, A., {Stubbs}, C.~W., {et~al.} 2008, \apj, 678, 865

\bibitem[{{Moretti} {et~al.}(2009){Moretti}, {Dall'Ora}, {Ripepi},
  {Clementini}, {Di Fabrizio}, {Smith}, {DeLee}, {Kuehn}, {Catelan}, {Marconi},
  {Musella}, {Beers}, \& {Kinemuchi}}]{Moretti2009}
{Moretti}, M.~I., {Dall'Ora}, M., {Ripepi}, V., {et~al.} 2009, \apjl, 699, L125

\bibitem[{{Mu{\~n}oz} {et~al.}(2012){Mu{\~n}oz}, {Geha}, {C{\^o}t{\'e}},
  {Vargas}, {Santana}, {Stetson}, {Simon}, \& {Djorgovski}}]{Munoz2012}
{Mu{\~n}oz}, R.~R., {Geha}, M., {C{\^o}t{\'e}}, P., {et~al.} 2012, \apjl, 753,
  L15

\bibitem[{{Musella} {et~al.}(2009){Musella}, {Ripepi}, {Clementini},
  {Dall'Ora}, {Kinemuchi}, {Fabrizio}, {Greco}, {Marconi}, {Smith}, {Radovich},
  \& {Beers}}]{Musella2009}
{Musella}, I., {Ripepi}, V., {Clementini}, G., {et~al.} 2009, \apjl, 695, L83

\bibitem[{{Musella} {et~al.}(2012){Musella}, {Ripepi}, {Marconi}, {Clementini},
  {Dall'Ora}, {Scowcroft}, {Moretti}, {Di Fabrizio}, {Greco}, {Coppola},
  {Bersier}, {Catelan}, {Grado}, {Limatola}, {Smith}, \&
  {Kinemuchi}}]{Musella2012}
{Musella}, I., {Ripepi}, V., {Marconi}, M., {et~al.} 2012, ArXiv e-prints

\bibitem[{{Nemec} {et~al.}(1988){Nemec}, {Wehlau}, \& {Mendes de
  Oliveira}}]{Nemec1988}
{Nemec}, J.~M., {Wehlau}, A., \& {Mendes de Oliveira}, C. 1988, \aj, 96, 528

\bibitem[{{Norris} {et~al.}(2010){Norris}, {Wyse}, {Gilmore}, {Yong}, {Frebel},
  {Wilkinson}, {Belokurov}, \& {Zucker}}]{Norris2010}
{Norris}, J.~E., {Wyse}, R.~F.~G., {Gilmore}, G., {et~al.} 2010, \apj, 723,
  1632

\bibitem[{{Oosterhoff}(1939)}]{Oosterhoff1939}
{Oosterhoff}, P.~T. 1939, The Observatory, 62, 104

\bibitem[{{Ripepi} {et~al.}(2012){Ripepi}, {Mancini}, {Cortecchia}, {Cusano},
  {Dall'Ora}, {Leccia}, {Molinaro}, {Penninpede}, {Clementini}, {Raiteri}, \&
  {Silvotti}}]{Ripepi2012}
{Ripepi}, V., {Mancini}, D., {Cortecchia}, F., {et~al.} 2012, Memorie della
  Societa Astronomica Italiana Supplementi, 19, 152

\bibitem[{{Sand} {et~al.}(2012){Sand}, {Strader}, {Willman}, {Zaritsky},
  {McLeod}, {Caldwell}, {Seth}, \& {Olszewski}}]{Sand2012}
{Sand}, D.~J., {Strader}, J., {Willman}, B., {et~al.} 2012, \apj, 756, 79

\bibitem[{{Sarajedini} {et~al.}(2006){Sarajedini}, {Barker}, {Geisler},
  {Harding}, \& {Schommer}}]{Sarajedini2006}
{Sarajedini}, A., {Barker}, M.~K., {Geisler}, D., {Harding}, P., \& {Schommer},
  R. 2006, \aj, 132, 1361

\bibitem[{{Schlafly} \& {Finkbeiner}(2011)}]{Schlafly2011}
{Schlafly}, E.~F., \& {Finkbeiner}, D.~P. 2011, \apj, 737, 103

\bibitem[{{Schlegel} {et~al.}(1998){Schlegel}, {Finkbeiner}, \&
  {Davis}}]{Schlegel1998}
{Schlegel}, D.~J., {Finkbeiner}, D.~P., \& {Davis}, M. 1998, \apj, 500, 525

\bibitem[{{Sesar} {et~al.}(2007){Sesar}, {Ivezi{\'c}}, {Lupton}, {Juri{\'c}},
  {Gunn}, {Knapp}, {DeLee}, {Smith}, {Miknaitis}, {Lin}, {Tucker}, {Doi},
  {Tanaka}, {Fukugita}, {Holtzman}, {Kent}, {Yanny}, {Schlegel}, {Finkbeiner},
  {Padmanabhan}, {Rockosi}, {Bond}, {Lee}, {Stoughton}, {Jester}, {Harris},
  {Harding}, {Brinkmann}, {Schneider}, {York}, {Richmond}, \& {Vanden
  Berk}}]{Sesar2007}
{Sesar}, B., {Ivezi{\'c}}, {\v Z}., {Lupton}, R.~H., {et~al.} 2007, \aj, 134,
  2236

\bibitem[{{Sesar} {et~al.}(2010){Sesar}, {Ivezi{\'c}}, {Grammer}, {Morgan},
  {Becker}, {Juri{\'c}}, {De Lee}, {Annis}, {Beers}, {Fan}, {Lupton}, {Gunn},
  {Knapp}, {Jiang}, {Jester}, {Johnston}, \& {Lampeitl}}]{Sesar2010}
{Sesar}, B., {Ivezi{\'c}}, {\v Z}., {Grammer}, S.~H., {et~al.} 2010, \apj, 708,
  717

\bibitem[{{Siegel}(2006)}]{Siegel2006}
{Siegel}, M.~H. 2006, \apjl, 649, L83

\bibitem[{{Siegel} \& {Majewski}(2000)}]{Siegel2000}
{Siegel}, M.~H., \& {Majewski}, S.~R. 2000, \aj, 120, 284

\bibitem[{{Simon} {et~al.}(2011){Simon}, {Geha}, {Minor}, {Martinez}, {Kirby},
  {Bullock}, {Kaplinghat}, {Strigari}, {Willman}, {Choi}, {Tollerud}, \&
  {Wolf}}]{Simon2011}
{Simon}, J.~D., {Geha}, M., {Minor}, Q.~E., {et~al.} 2011, \apj, 733, 46

\bibitem[{{Smith}(1995)}]{Smith1995}
{Smith}, H.~A. 1995, Cambridge Astrophysics Series, 27

\bibitem[{{Smith} {et~al.}(2009){Smith}, {Catelan}, \&
  {Clementini}}]{Smith2009}
{Smith}, H.~A., {Catelan}, M., \& {Clementini}, G. 2009, in American Institute
  of Physics Conference Series, Vol. 1170, American Institute of Physics
  Conference Series, ed. J.~A. {Guzik} \& P.~A. {Bradley}, 179--187

\bibitem[{{Stetson}(1987)}]{Stetson1987}
{Stetson}, P.~B. 1987, \pasp, 99, 191

\bibitem[{{Stetson}(1994)}]{Stetson1994}
---. 1994, \pasp, 106, 250

\bibitem[{{Vivas} {et~al.}(2004){Vivas}, {Zinn}, {Abad}, {Andrews}, {Bailyn},
  {Baltay}, {Bongiovanni}, {Brice{\~n}o}, {Bruzual}, {Coppi}, {Della Prugna},
  {Ellman}, {Ferr{\'{\i}}n}, {Gebhard}, {Girard}, {Hernandez}, {Herrera},
  {Honeycutt}, {Magris}, {Mufson}, {Musser}, {Naranjo}, {Rabinowitz},
  {Rengstorf}, {Rosenzweig}, {S{\'a}nchez}, {S{\'a}nchez}, {Schaefer},
  {Schenner}, {Snyder}, {Sofia}, {Stock}, {van Altena}, {Vicente}, \&
  {Vieira}}]{Vivas2004}
{Vivas}, A.~K., {Zinn}, R., {Abad}, C., {et~al.} 2004, \aj, 127, 1158

\bibitem[{{Walsh} {et~al.}(2007){Walsh}, {Jerjen}, \& {Willman}}]{Walsh2007}
{Walsh}, S.~M., {Jerjen}, H., \& {Willman}, B. 2007, \apjl, 662, L83

\bibitem[{{Willman} {et~al.}(2011){Willman}, {Geha}, {Strader}, {Strigari},
  {Simon}, {Kirby}, {Ho}, \& {Warres}}]{Willman2011}
{Willman}, B., {Geha}, M., {Strader}, J., {et~al.} 2011, \aj, 142, 128

\bibitem[{{Willman} \& {Strader}(2012)}]{Willman2012}
{Willman}, B., \& {Strader}, J. 2012, \aj, 144, 76

\bibitem[{{Willman} {et~al.}(2005{\natexlab{a}}){Willman}, {Blanton}, {West},
  {Dalcanton}, {Hogg}, {Schneider}, {Wherry}, {Yanny}, \&
  {Brinkmann}}]{Willman2005a}
{Willman}, B., {Blanton}, M.~R., {West}, A.~A., {et~al.} 2005{\natexlab{a}},
  \aj, 129, 2692

\bibitem[{{Willman} {et~al.}(2005{\natexlab{b}}){Willman}, {Dalcanton},
  {Martinez-Delgado}, {West}, {Blanton}, {Hogg}, {Barentine}, {Brewington},
  {Harvanek}, {Kleinman}, {Krzesinski}, {Long}, {Neilsen}, {Nitta}, \&
  {Snedden}}]{Willman2005b}
{Willman}, B., {Dalcanton}, J.~J., {Martinez-Delgado}, D., {et~al.}
  2005{\natexlab{b}}, \apjl, 626, L85

\bibitem[{{Wolf} {et~al.}(2010){Wolf}, {Martinez}, {Bullock}, {Kaplinghat},
  {Geha}, {Mu{\~n}oz}, {Simon}, \& {Avedo}}]{Wolf2010}
{Wolf}, J., {Martinez}, G.~D., {Bullock}, J.~S., {et~al.} 2010, \mnras, 406,
  1220

\bibitem[{{Zucker} {et~al.}(2006{\natexlab{a}}){Zucker}, {Belokurov}, {Evans},
  {Kleyna}, {Irwin}, {Wilkinson}, {Fellhauer}, {Bramich}, {Gilmore}, {Newberg},
  {Yanny}, {Smith}, {Hewett}, {Bell}, {Rix}, {Gnedin}, {Vidrih}, {Wyse},
  {Willman}, {Grebel}, {Schneider}, {Beers}, {Kniazev}, {Barentine},
  {Brewington}, {Brinkmann}, {Harvanek}, {Kleinman}, {Krzesinski}, {Long},
  {Nitta}, \& {Snedden}}]{Zucker2006b}
{Zucker}, D.~B., {Belokurov}, V., {Evans}, N.~W., {et~al.} 2006{\natexlab{a}},
  \apjl, 650, L41

\bibitem[{{Zucker} {et~al.}(2006{\natexlab{b}}){Zucker}, {Belokurov}, {Evans},
  {Wilkinson}, {Irwin}, {Sivarani}, {Hodgkin}, {Bramich}, {Irwin}, {Gilmore},
  {Willman}, {Vidrih}, {Fellhauer}, {Hewett}, {Beers}, {Bell}, {Grebel},
  {Schneider}, {Newberg}, {Wyse}, {Rockosi}, {Yanny}, {Lupton}, {Smith},
  {Barentine}, {Brewington}, {Brinkmann}, {Harvanek}, {Kleinman}, {Krzesinski},
  {Long}, {Nitta}, \& {Snedden}}]{Zucker2006a}
---. 2006{\natexlab{b}}, \apjl, 643, L103

\end{thebibliography}

\clearpage

\begin{deluxetable}{lccccccc}
\tabletypesize{\scriptsize}
\tablecaption{$\langle [Fe/H] \rangle$ and $\langle P_{ab} \rangle$ for Milky Way Dwarf Companions}
\tablehead{
\colhead{Object} &
\colhead{$N_{RRab}$\tablenotemark{a}} &
\colhead{$\langle [Fe/H] \rangle$} &
\colhead{$\sigma_{\langle [Fe/H] \rangle}$} &
\colhead{$\langle P_{ab} \rangle$} &
\colhead{$\sigma_{\langle P_{ab} \rangle}$} &
\colhead{$\langle P_{ab} \rangle$ ref} &
\colhead{$\langle [Fe/H] \rangle$ ref}
}
\startdata
Cvn I & 18 & -1.962 & 0.038 & 0.600 & 0.006 & \cite{Kuehn2008} & WS12, K10 \\
Herc & 6 & -2.518 & 0.140 & 0.678 & 0.013 & \cite{Musella2012} & WS12, K08\\
For & 396 & -1.025 & 0.012 & 0.585 & 0.002 & \cite*{Bersier2002} & WS12, K10\\
Dra & 123 & -1.946 & 0.024 & 0.619 & 0.004 & \cite{Bonanos2004} & WS12, K10\\
Leo IV & 3 & -2.363 & 0.230 & 0.655 & 0.028 & \cite{Moretti2009} & WS12, K08\\
Sex & 26 & -1.966 & 0.039 & 0.606 & 0.010 & \cite{Mateo1995} & WS12, K10\\
Leo I & 47 & -1.450 & 0.011 & 0.602 & 0.009 & \cite{Held2001} & WS12, K10\\
Leo II & 103 & -1.670 & 0.024 & 0.620 & 0.006 & \cite*{Siegel2000} & WS12, K10\\
UMi & 47 & -2.112 & 0.027 & 0.638 & 0.009 & \cite{Nemec1988} & WS12, K10\\
Scl & 129 & -1.726 & 0.024 & 0.584 & 0.007 & \cite{Kaluzny1995} & WS12, K10\\
Boo I & 7 & -2.531 & 0.132 & 0.691 & 0.034 & \cite{Siegel2006} & \cite{Norris2010}\\
ComBer & 1 & -2.640 & 0.100 & 0.670 & \nodata & \cite{Musella2009} & WS12, K08\\
Cvn II & 1 & -2.444 & 0.178 & 0.743 & \nodata & \cite{Greco2008} & WS12, K08\\
UMa I & 5 & -2.334 & 0.128 & 0.628 & 0.032 & \cite{Garofalo2013} & WS12, K08\\
UMa II & 1 & -2.357 & 0.204 & 0.659 & \nodata  & \cite{Dall'Ora2012} & WS12, K08\\
Seg2 & 1 & -2.257 & 0.140 & 0.748 & \nodata & present work & \cite{Kirby2013}\tablenotemark{b}\\
\enddata
\tablecomments{WS12 = \cite{Willman2012}; K08 = \cite{Kirby2008}; K10 = \cite{Kirby2010}. When two references are listed for $\langle [Fe/H] \rangle$, the WS12 reference contains the calculated average and uncertainty and the K08 or K10 reference contains the original [Fe/H] measurements.}
\tablenotetext{a}{As the cited RRL surveys are not necessarily complete for the more luminous dwarfs, the number of RRab stars cited as belonging to these dwarfs may be underestimated. Additionally, RRab stars with abnormal or uncertain classifications were not included in the total count.}
\tablenotetext{b}{The mean metallicity of Segue 2 was calculated using the technique described in \citet{Willman2012} from the individual metallicities published in \citet{Kirby2013}.}
\end{deluxetable}

\end{document}